\documentclass[prb,twocolumn,showpacs,preprintnumbers,amsmath,amssymb]{revtex4}
\usepackage{graphicx}% Include figure files
\usepackage{dcolumn}% Align table columns on decimal point
\usepackage{bm}% bold math
\usepackage{amssymb}
\usepackage{epstopdf}
\newcommand{\lb}{\left[}
\newcommand{\rb}{\right]}
\newcommand{\lp}{\left(}
\newcommand{\rp}{\right)}

\newcommand{\be}{\begin{equation}}
\newcommand{\ee}{\end{equation}}
\newcommand{\Tr}{\,{\rm Tr}\,}

\renewcommand{\Im}{\,{\rm Im}\,}
\renewcommand{\phi}{\varphi}
\renewcommand{\epsilon}{\varepsilon}
\renewcommand{\vec}[1]{{\bf #1}}

\begin{document}
\title{Electron Interactions in Bilayer Graphene: Marginal Fermi Liquid Behavior and Zero-Bias Anomaly}
\author{Rahul Nandkishore and Leonid Levitov}
\affiliation{Department of Physics, Massachusetts Institute of Technology, 77 Massachusetts Avenue, Cambridge MA02139}

\begin{abstract}
We analyze the many-body properties of bilayer graphene (BLG) at charge neutrality, governed by long range interactions between electrons. Perturbation theory in a large number of flavors is used in which the interactions are described within a random phase approximation, taking account of dynamical screening effect. Crucially, the dynamically screened interaction retains some long range character, resulting in $\log^2$ renormalization of key quantities. We carry out the perturbative renormalization group calculations to one loop order, and find that BLG behaves to leading order as a marginal Fermi liquid. Interactions produce a log squared renormalization of the quasiparticle residue and the interaction vertex function, while all other quantities renormalize only logarithmically. We solve the RG flow equations for the Green function with logarithmic accuracy, and find that the quasiparticle residue flows to zero under RG. At the same time, the gauge-invariant quantities, such as the compressibility, remain finite to $\log^2$ order, with subleading logarithmic corrections. The key experimental signature of this marginal Fermi liquid behavior is a strong suppression of the tunneling density of states, which manifests itself as a zero bias anomaly in tunneling experiments in a regime where the compressibility is essentially unchanged from the non-interacting value. 
\end{abstract}

\maketitle
\section{Introduction}

Bilayer graphene (BLG), due to its unique electronic structure of a two dimensional gapless semiconductor with quadratic dispersion \cite{Novoselov}, offers an entirely new setting for investigating many body phenomena. In sharp contrast to single layer graphene, the density of states in BLG does not vanish at charge neutrality, and thus even arbitrarily weak interactions can trigger phase transitions. Theory predicts instabilities to numerous strongly correlated gapped and gapless states in BLG \cite{Min, Nandkishore, Zhang, Kivelson, Vafek}. 
%\mpar{Are these the earliest predictions of instability in BLG?}
These instabilities have been analyzed in models with unscreened long-range interactions \cite{Min}, dynamically screened long-range interactions \cite{Nandkishore} and in models where the interactions are treated as short range \cite{Kivelson, Zhang, Vafek}. Irrespective of the form of the interaction, the instability develops only logarithmically with the energy scale. However, dynamically screened Coulomb interactions have been shown to produce $\log^2$ renormalization of the self energy\cite{Yang} and vertex function \cite{Nandkishore}. Such strong renormalization can result in significant departures from non-interacting behavior on energy scales much greater than those characteristic for the onset of gapped states.  
% the critical energy scale for the onset of gapped states. 
However, there is as yet no systematic treatment of the $\log^2$ divergences. In this paper, we provide a systematic treatment of the effects of dynamically screened Coulomb interactions, focusing on the renormalisation of the Green function, and using the framework of the perturbative renormalization group (RG). 

% We calculate the RG flow for BLG
We analyze the RG flow perturbatively in the number of flavors, given by $N=4$ in BLG. 
% to with long range dynamically screened Coulomb interactions. We work perturbatively 
We use perturbation theory developed about the non-interacting fixed point, and calculate the renormalization of the fermion Green function and of the Coulomb interactions. We demonstrate that the quasiparticle residue and the Coulomb vertex function undergo $\log^2$ renormalization, while all other quantities renormalize only logarithmically. The quasiparticle residue and the Coulomb vertex function, moreover, are not independent, but are related by a Ward identity which stems from gauge invariance symmetry. %\mpar{drop particle number conservation; this is really just the old good gauge invariance}
Therefore, at $\log^2$ order, BLG behaves as a marginal Fermi liquid. 

We solve the RG flow equations with logarithmic accuracy, 
% and discover 
finding that the quasiparticle residue flows to zero under RG. This behavior manifests itself in a zero bias anomaly in the tunneling density of states (TDOS). We conclude by extracting the sub-leading (single log) renormalization of the electron mass, as a correction to the log square RG. This calculation allows us to predict the interaction renormalization of the electronic compressibility in BLG, a quantity which is interesting both because it is directly experimentally measurable, and because it allows us to contrast the slow single log renormalization of the compressibility with the fast $\log^2$ renormalization of the TDOS. %Our work differs from earlier RG based investigations of BLG \cite{Zhang, Vafek} in that we work with dynamically screened long range interactions, whereas the works \cite{Zhang, Vafek} assumed interactions were short range. The physics described in this paper only manifests itself when we work with dynamically screened long range interactions. 

The structure of the perturbative RG for BLG has strong similarities to the 
%LL well known 
perturbative RG treatment of the one dimensional Luttinger liquids \cite{Zhang, DzyaloshinskiiLarkin,Shankar,Giamarchi}. We recall that in the Luttinger liquids, the Green function acquires an anomalous scaling dimension, which manifests itself in a power law behaviour of a quasiparticle residue that vanishes on shell. 
% Meanwhile, in the Luttinger liquids, 
% as in BLG, 
%LL Second, 
In addition, the electronic compressibility in the Luttinger liquids remains finite even as the quasiparticle residue flows to zero. Finally, in the Luttinger liquids, there are logarithmic divergences in Feynman diagrams describing scattering in the particle-particle and particle hole channels, corresponding to mean field instabilities to both Cooper pairing and charge density wave ordering. However, when both instabilities are taken into account simultaneously within the framework of the RG, they cancel each other out, so that there is no instability to any long range ordered phase at low energies \cite{DzyaloshinskiiLarkin}. 

Exactly the same behavior follows from our RG analysis of BLG, including the cancellation of the vertices responsible for the pairing and charge density ordering.
% also manifests itself in BLG. 
However, the diagrams in this instance are $\log^2$ divergent, and even after the leading $\log^2$ divergences are canceled out, there remains a subleading single log instability. Nevertheless, this single log instability manifests itself on much lower energy scales than the $\log^2$ RG flow. 
Therefore, over a large range of energies, bilayer graphene can be viewed as a two dimensional analogue of the one dimensional Luttinger liquids. 

Our treatment of the $\log^2$ renormalization in BLG 
% presented below
% in this work
% renormalization of the Green function and the vertex function of a $\log^2$ form 
is somewhat reminiscent of the situation arising in two-dimensional disordered metals\cite{Aronov}. 
% Similar to the electron system in BLG, discussed below, the $\log^2$ divergences 
In the latter, the $\log^2$  divergences of the Green function and of the vertex function
%arise in the analysis of Ref.[\onlinecite{Aronov}] due to 
stem from the properties of dynamically screened Coulomb interactions, which exhibit ``unscreening" for the transferred frequencies and momenta such that $\omega/q^2$ is large compared to the diffusion coefficient. Furthermore, the divergent corrections to the Fermi-liquid parameters, as well as conductivity, compressibility and other two-particle quantities in these systems, are only logarithmic. This allows to describe the RG flow of the Green function
% and vertex function 
due to the $\log^2$ divergences by a single RG equation \cite{Finkelstein} of the form 
\begin{equation}
\partial G/\partial\xi=-\frac{\xi}{4\pi^2 g}G,
\end{equation}
where $g$ is the dimensionless conductance. The suppression of the quasiparticle residue, described by this equation, manifests itself in a zero-bias anomaly in the tunneling density of states, readily observable by transport measurements.
% in disordered metals.
%\section{Self-consistent one loop perturbative RG}

\section{Dynamically screened interaction}

We begin by reviewing some basic facts about BLG. BLG consists of two AB stacked graphene sheets (Bernal stacking). The low-energy Hamiltonian can be described in a `two band' approximation, neglecting the higher bands that are separated from the Dirac point by an energy gap $W \sim 0.4$ eV \cite{Novoselov}. There is four-fold spin/valley degeneracy. The wavefunction of the low energy electron states 
% exist 
resides on the $A$ sublattice of one layer and the B sublattice of the other layer.
% , and there are two spins and two valleys. 
The non-interacting spectrum consists of quadratically dispersing quasiparticle bands $E_{\pm} = \pm p^2/2m$ with band mass $m \approx 0.054 m_e$. We work throughout at charge neutrality, when the Fermi surface consists of Fermi points. The discrete 
point-like nature of the Fermi surface is responsible for most of the similarities to the Luttinger liquids. 

Although the canonical Hamiltonian has opposite chirality in the two valleys, a suitable unitary transformation on the spin-valley-sublatttice space brings the Hamiltonian to a form where there are four flavors of fermions, each governed by the same $2\times2$ 
quadratic Dirac-type Hamiltonian \cite{Nandkishore2}. We introduce the Pauli matrices that act on the sublattice space $\tau_i$, and define $\tau_{\pm} = \tau_1 \pm i \tau_2$, and $p_{\pm} = p_x \pm i p_y$, and hence write\footnotemark
\begin {eqnarray}
 \label{eq: Hamiltonian}
&& H = 
% \sum_{\rm{p}, \alpha}\psi^\dagger_\alpha(\vec p) H^{\alpha}_0(\vec{p})\psi_\alpha(\vec p)
H_0+ \frac{e^2}{2\kappa}\sum_{\rm x,x'}\frac{n(\vec x)n(\vec x')}{|\vec x-\vec x'|}
,
% n(\vec x)V(\vec{x}, \vec{x'})n(\vec x') 
\\
&& H_0 = \sum_{\vec p,\sigma}\psi^\dagger_{\vec p,\sigma}\lp \frac{p_+^2}{2m} \tau_+ + \frac{p_-^2}{2m} \tau_-\rp \psi_{\vec p,\sigma}
.
% \frac{p_+^2\tau_+ +p_-^2\tau_-}{2m}\psi_{\vec p,\alpha}
% \frac{p_+^2}{2m} \tau_+ + \frac{p_-^2}{2m} \tau_-
% ,  \quad
% V(\vec{x}, \vec{x'}) =  \frac{e^2}{\kappa|\vec x-\vec x'|} \nonumber\\
%n(\vec x)&=&\sum_\alpha\psi_\alpha^\dagger(\vec x)\psi_\alpha(\vec x)\nonumber.
 \end{eqnarray}
Here $\sigma=1,2,3,4$ is a flavour index, $n(\vec{x})=\sum_\sigma n_\sigma(\vec{x})$ is the electron density, summed over spins, valleys and sublattices, while the dielectric constant $\kappa$ incorporates the effect of polarization of the substrate. Note that the single-particle Hamiltonian $H_0$ takes the same form for each of the four fermion flavors, and is thus $SU(4)$ invariant under unitary rotations in the flavor space. 

\footnotetext{We have performed a unitary transformation on the Hamiltonian, as outlined in Ref.[\onlinecite{Nandkishore2}], to clearly manifest the symmetries. As a consequence, our `valley' and `sublattice' variables are not the physical valley and sublattice variables, but are linear combinations thereof. }

The Coulomb interaction sets a characteristic length scale and a  characteristic energy scale (``Bohr radius and Rydberg energy'')\cite{Nandkishore}:
% $E_0$ and a characteristic length scale $a_0$ respectively \cite{Nandkishore}, where 
%
\begin{equation}
a_0 = \frac{\hbar^2 \kappa}{ me^2} \approx 10\kappa  \,{\rm \AA}
, \qquad 
E_0 = \frac{e^2}{\kappa a_0} \approx \frac{1.47}{\kappa^2} \,\rm{eV} 
% \kappa^{-2} 
.
\label{eq: scales}
\end{equation}
%
%The Hamiltonian Eq.\ref{eq: Hamiltonian} is invariant under SU(4) rotations in spin-valley space \cite{Levitov}, a symmetry that will be preserved by the renormalization group procedure. This symmetry is an artefact of the approximations leading to Eq.\ref{eq: Hamiltonian}, 
%
In Eq.(\ref{eq: Hamiltonian}), we have approximated by assuming that the interlayer and intra-layer interaction are equal. This approximation may be justified by noting that the interlayer spacing $d \approx 3\,{\rm \AA}$ is much less than the characteristic lengthscale $a_0$,
% set by the interaction (
Eq.(\ref{eq: scales}). Within this approximation, the Hamiltonian (\ref{eq: Hamiltonian}) is invariant under $SU(4)$ flavour rotations
%, and under time reversal 
\cite{Nandkishore2}. 
%\mpar{time reversal is an exact symmetry; what did it mean to say?}

We note that for $\kappa\sim 1$ the energy $E_0$
% (Eq.\ref{eq: scales}) 
value is comparable to the energy gap parameter $W \sim 0.4\,{\rm eV}$ of the higher BLG bands (see Ref.[\onlinecite{McCann}] for a discussion of four band model of BLG). %describing the states antihybridized by interlayer tunnel coupling.
% of BLG. 
% 
% \mpar{OK?}
This suggests that there is some interaction induced mixing with the higher bands of BLG. However, since a four band analysis is exceedingly tedious, 
% we develop our analysis in 
here we focus on the weak coupling limit $E_0 \ll W$, where the two band approximation, Eq.(\ref{eq: Hamiltonian}), is rigorously accurate. We perform all our calculations in this weak coupling regime, and then 
% perform an analytic continuation 
extrapolate the result to $E_0 \approx 1.47\,{\rm eV} \kappa^{-2}$. Since the low energy properties should be independent of the higher bands, we believe this approximation correctly captures, at least qualitatively, the essential physics in BLG. Meanwhile, since $W$ is the maximum energy scale up to which the two band Hamiltonian, Eq.(\ref{eq: Hamiltonian}), is valid, we use $W$ as the initial UV cutoff for our RG analysis. %\mpar{OK?}%The effect of interband mixing will be further discussed below, following Eq.(\ref{eq: bandwidth}). 

We wish to obtain a RG flow for the problem (\ref{eq: Hamiltonian})
% Hamiltonian Eq.\ref{eq: Hamiltonian} 
by systematically integrating out the high energy modes. However, the implementation of this strategy is complicated by the long range nature of the unscreened Coulomb interaction.
% in Eq.\ref{eq: Hamiltonian}. 
Within perturbation theory, the long range interaction gives contributions which are relevant at tree level, 
% which makes it very 
making it difficult to 
% obtain 
come up with a meaningful perturbative RG scheme. Therefore, it is technically convenient to perform a two-step calculation, where we first take into account screening within the random-phase approximation (RPA), and then carry out an RG calculation with the RPA screened effective interaction. We emphasize that it is necessary to consider the full \emph{dynamic} RPA screening of the Coulomb interaction, since a static screening approximation does not capture the effects we discuss below. 

The dynamically screened interaction may be calculated by summing over the RPA series of bubble diagrams, to obtain a screened interaction. The RPA approach to screening may be justified by invoking the large number $N=4$ of fermion species in BLG. The screened interaction takes the form 
%LL in Fourier space takes the form 
\begin{eqnarray}
\label{eq: Ueff}
U(\omega, \vec{q}) = \frac{2\pi e^2 }{\kappa |\vec q| - 2\pi e^2 \Pi(\omega, \vec q)}
.
\end{eqnarray}
Here $\Pi(\omega, \vec q)$ is the non-interacting polarization function, which can be evaluated analytically\cite{Nilsson,Nandkishore}. Here we will need an expression for $\Pi(\omega, \vec{q})$ in terms of Matsubara frequencies $\omega$, derived in Ref.[\onlinecite{Nandkishore}], where it was shown that the quantity $\Pi(\omega, \vec q)$ 
% The non-interacting polarization function was calculated analytically in \cite{Nandkishore}. It was shown that $\Pi$ 
depends on a single parameter $2m\omega/q^2$, and is well described by the approximate form
% \cite{Nandkishore}
\begin{eqnarray}\label{eq: polfn}
\Pi(\omega, \vec{q}) &=& - \frac{N m}{2\pi} \frac{\ln4\, \frac{\vec q^2}{2m}}{\sqrt{\big(\frac{\vec q^2}{2m}\big)^2+ u \omega^2}} 
,\quad
u = \frac{4\ln^24}{\pi^2}
,
\end{eqnarray}
% where $u = (2\ln4/\pi)^2$ and 
where $N=4$ is the number of fermion species. 
% This approximate form 
% 
The dependence (\ref{eq: polfn}) reproduces $\Pi(\omega, \vec{q})$ exactly in the limits $\omega\ll {\vec q}^2/2m $ and $\omega\gg {\vec q}^2/2m $, and interpolates accurately in between. We discover upon substituting Eq.(\ref{eq: polfn}) in Eq.(\ref{eq: Ueff}) that the dynamically screened interaction is retarded in time, but crucially is only marginal at tree level. It therefore becomes possible to develop the RG analysis perturbatively in weak coupling strength, by taking the limit of $N\gg 1$.
% , before finally performing an analytic continuation to $N=4$. Note that 

Since the quantity $\Pi(\omega, \vec{q})$ vanishes when $q \rightarrow 0$, the RPA screened interaction (\ref{eq: Ueff}) retains some long range character, exhibiting ``unscreening'' for $\omega\gg \vec q^2/2m$. This will lead to 
%LL unusually strong 
divergences in Feynman diagrams of a $\log^2$ character. 

\section{Setting up the RG}
\label{sec: setup}
To calculate the RG flow of the Hamiltonian, Eq.(\ref{eq: Hamiltonian}), in the weak coupling regime, %it is convenient to follow \cite{Gonzalez} and to replace the four fermion Coulomb interaction by an interaction with an auxiliary scalar field $\phi$.
we begin by writing the zero-temperature partition function $\Phi$ as an imaginary-time functional field integral. We have
\begin{eqnarray}
&& \Phi = \int D \psi^{\dag} D\psi \exp\left(-S_0[\psi^{\dag}, \psi] - S_1[\psi^{\dag}, \psi] \right), \label{eq: partition function}  \\
&& S_0 = \sum_{\sigma} \int \frac{d\omega d^2p}{(2\pi)^3} \psi^{\dag}_{\sigma, \omega, \vec{p}}\!\left(\frac{-i\omega + H^{\sigma}_0(\vec{p})}{Z}\right)\!\psi_{\sigma, \omega, \vec{p}}, \label{eq: bare action}\\
&& S_1 = \frac{1}{2} \int \frac{d\omega d^2p}{(2\pi)^3} \Gamma^2 U(\omega, \vec{q}) n_{\omega, \vec{q}}n_{-\omega, -\vec{q}}  + S_2 . \label{eq: coupling}
\end{eqnarray}
Here the $\psi$ fields are Grassman valued (fermionic) fields with flavour (spin-valley) index $\sigma$, while $\omega$ is a fermionic Matsubara frequency, $\Gamma$ is a vertex renormalization parameter,
%\mpar{a coupling constant or a vertex renormalization parameter?} 
$Z$ is the quasiparticle residue, and $n_{\omega, \vec{q}}$ is the Fourier transform of the electron density, summed over spins, valleys and sublattices. The effective interaction $U(\omega, \vec{q})$ is given by Eq.(\ref{eq: Ueff}). 
% Meanwhile, the term $S_2$ represents 
The term $S_2$ is included tentatively to represent more complicated interactions that may be generated under RG. In the 
% initial 
bare theory, $\Gamma = 1$, $Z=1$ and $S_2 = 0$. The theory is defined with the initial UV cutoff $\Lambda_0$. Since the two band model, Eq.\ref{eq: Hamiltonian}, is only justified on energy scales less than the gap $W \approx 0.4 eV$ to the higher bands in BLG, we conservatively identify $\Lambda_0 = W$. Our main results will be independent of $\Lambda_0$. %\mpar{OK?}

% We also note that 
As we shall see, the RG flow will inherit the symmetries of the Hamiltonian, Eq.(\ref{eq: Hamiltonian}), strongly constraining the possible terms $S_2$. The relevant symmetries are particle-hole symmetry, time reversal symmetry, SU(4) flavour symmetry \cite{Nandkishore2}, and the symmetry of the Hamiltonian under the transformation $e^{i\theta\tau_3}R(\theta/2)$, where $R(\theta)$ generates spatial rotations, $R(\theta) p_\pm=e^{\pm i\theta}p_\pm$.
%  in the plane. 
% \mpar{OK?}
 %The auxiliary field $\phi$ introduced in Eq.\ref{eq: partition function} has the propagator
%\begin{equation}
%i \langle T \phi(\vec{r},t) \phi(\vec{r}', t')\rangle = \frac{2\pi e^2 }{\kappa q - 2\pi e^2 \Pi(\omega, q)} \label{eq: dressed propagator}
%\end{equation}

We will employ an RG scheme which treats frequency $\omega$ on the same footing as $p^2/2m$, in order to preserve the form of the free action Eq.(\ref{eq: bare action}) under RG. Thus, we integrate out the shell of highest energy fermion modes 
\begin{equation}
\Lambda' < \sqrt{\omega^2 + \lp \frac{{\vec p}^2}{2m}\rp^2} < \Lambda
, \label{eq: shell}
\end{equation}
and subsequently rescale $\omega \to \omega (\Lambda/\Lambda')$, $p \to p (\Lambda/\Lambda')^{1/z}$,
% $p \rightarrow p (\sqrt{\Lambda_0/\Lambda_1})$ and $\omega \rightarrow \omega (\sqrt{\Lambda_0/\Lambda_1})^z$, 
%
% \be
% \omega \rightarrow \omega (\Lambda_0/\Lambda_1),\quad
% p \rightarrow p (\Lambda_0/\Lambda_1)^{1/z}
% ,
% \ee
%
where $z$ is the dynamical critical exponent \cite{Shankar}, which takes value $z=2$ at tree level. Because the value $z=2$ is not protected by any symmetry, it may acquire renormalization corrections.
%  as a result of be corrections to the tree level value of $z=2$, \mpar{OK?}however, 
However, it will follow from our analysis that the quasiparticle spectrum does not renormalize at leading $\log^2$  order, so that the 
%LL dynamic critical 
exponent $z$ does not flow at leading order. We therefore use $z=2$ for the rest of the paper, which corresponds to scaling dimensions $[\omega] = 1$ and $[p^2] = 1$. 
%\mpar{the values $[\omega]=1$ and $[p^2]=1$ agree with our choice of RG time}
%\mpar{ explain the dyn cri exp and/or cite some source}
 Under such an RG transformation, 
% the terms in Eq.\ref{eq: bare action},\ref{eq: coupling} 
the Lagrangian density in momentum space has scaling dimension $[\mathcal{L}]=2$, and we have tree level scaling dimensions $[\psi]=1/2$ and $[\Gamma]=[Z] = 0$ respectively. 

Given these tree level scaling dimension values, it can be seen that all potentially relevant terms arising as part of $S_2$ must involve four fermion fields. 
% These tree level scaling dimensions place strong constraints on the term $S_2$ in Eq.\ref{eq: coupling}. Any 
Indeed, any term involving more than four $\psi$ fields will be irrelevant at tree level under RG, and may be neglected. The terms with odd numbers of $\psi$ fields are forbidden by charge conservation, while the quadratic terms $\Delta_{ij} \psi_i^{\dag} \psi_j$ cannot be generated under perturbative RG, since they break the symmetries of the Hamiltonian listed above\footnotemark.
% \mpar{OK?}. %since it breaks either particle-hole symmetry, time reversal symmetry, SU(4) flavour symmetry \cite{Nandkishore2}, or the symmetry of the Hamiltonian under the transformation $e^{i\theta}R(\theta/2)$, where $R(\theta)$ generates a spatial rotation by $\theta$ in the plane. 
\footnotetext{The symmetry of the Hamiltonian may be spontaneously broken. However, the energy scale for spontaneous symmetry breaking is set by the subleading single log flows \cite{Nandkishore} and is lower than the energy scale for the phenomena discussed in this paper. }
Thus, the only potentially relevant terms that could arise under perturbative RG take the form of a four point interaction which may be written as
\begin{equation}
S_2 = \frac{1}{2}\int d^3x d^3x' \Upsilon^{\sigma \sigma'}_{ijkl} \psi^{\dag}_{\sigma, i}(x) \psi_{\sigma, j}(x) \psi^{\dag}_{\sigma', k}(x') \psi_{\sigma', l}(x')
, \label{eq: four point vertex}
\end{equation}
where $x=(\vec r, t)$, $x'=(\vec r', t')$, 
Here $\Upsilon$ 
%LL may be interpreted as 
is an effective four particle vertex, which is marginal at tree level, the indices $\sigma, \sigma'$ refer to the flavour (spin-valley) of the interacting particles, and $i,j,k,l$ are sublattice indices. 

The symmetries of the Hamiltonian, Eq.(\ref{eq: Hamiltonian}), impose strong constraints on the spin-valley-sublattice structure of the four point vertex $\Upsilon$. Since the Coulomb interaction does not change fermion flavour (spin or valley), and the electron Green function is diagonal in flavour space, the vertex $\Upsilon$ cannot change fermion flavour. Moreover, the $SU(4)$ flavour symmetry of the Hamiltonian implies that $\Upsilon$ does not depend on the flavour index of the interacting particles, and we may therefore drop the indices $\sigma, \sigma'$ in Eq.(\ref{eq: four point vertex}). Finally, the bare Hamiltonian (\ref{eq: Hamiltonian}) is invariant under combined pseudospin/spatial rotations through $e^{i\theta \tau_3} R(\theta/2)$.
% the action of the operator $e^{i\theta \tau_3} R(\theta/2)$, where $R(\theta)$ generates a spatial rotation by $\theta$ in the plane. 
%\mpar{show the sublattice structure on a figure}
This symmetry further restricts the form of four point vertices in Eq.(\ref{eq: four point vertex}) to have sublattice structure $\Upsilon_{iijj}$ or $\Upsilon_{ijji}$ only \footnotemark. That is, the allowed scattering processes are restricted to $(AA) \rightarrow (AA)$, $(AB) \rightarrow (AB)$ and $(AB) \rightarrow (BA)$. We note that the processes $(AB) \rightarrow (AB)$ and $(AB) \rightarrow (BA)$ are distinct, since the particles have flavour, and the interaction (\ref{eq: Ueff}) is not short range.
\footnotetext{In that, we ignore vertices of the form $\Upsilon_{AAAB}\partial_+^2$, $\Upsilon_{AABA}\partial_-^2$, and other similar terms, which are allowed by symmetries, but are irrelevant in the RG sense.}
%We further note that the symmetry of the Hamiltonian Eq.\ref{eq: Hamiltonian} under simultaneous sublattice interchange and parity inversion 

Below we obtain the RG flow for bilayer graphene, working in the manner of Ref.[\onlinecite{Shankar}]. We consider the partition function, Eq.(\ref{eq: partition function}), where the interaction is given by Eq.(\ref{eq: Ueff}). Starting from this action, supplied with ultraviolet (UV) cutoff $\Lambda_0$, we systematically integrate out the shell of highest energy fermion modes, Eq.(\ref{eq: shell}).
%
%\begin{equation}
%\Lambda_1 < \sqrt{\omega^2 + \big(\frac{p^2}{2m}\big)^2} < \Lambda_0. \label{eq: shell}
%\end{equation}
%
We perform the integrals perturbatively in the interaction, Eq.(\ref{eq: Ueff}). This corresponds to a perturbation theory in small $\Gamma^2 Z^2/N$. We carry out our calculations to one loop order, and examine the renormalization, in turn, of the electron Green function (Sec.\ref{sec4}), the vertex function
$\Gamma$ (Sec.\ref{sec5}) and the four point vertex $\Upsilon$ (Sec.\ref{sec6}).
% respectively. 

\section{Self-consistent renormalization of the electron Green function}
\label{sec4}

\begin{figure}
\large{a)} \includegraphics[scale = 0.21]{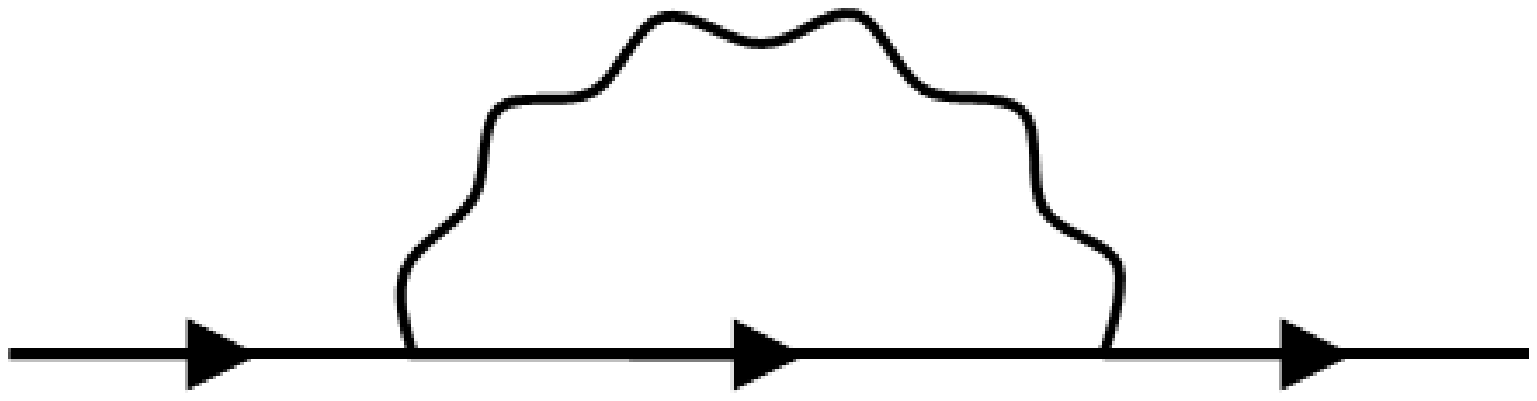}
\large{b)}\includegraphics[scale = 0.21]{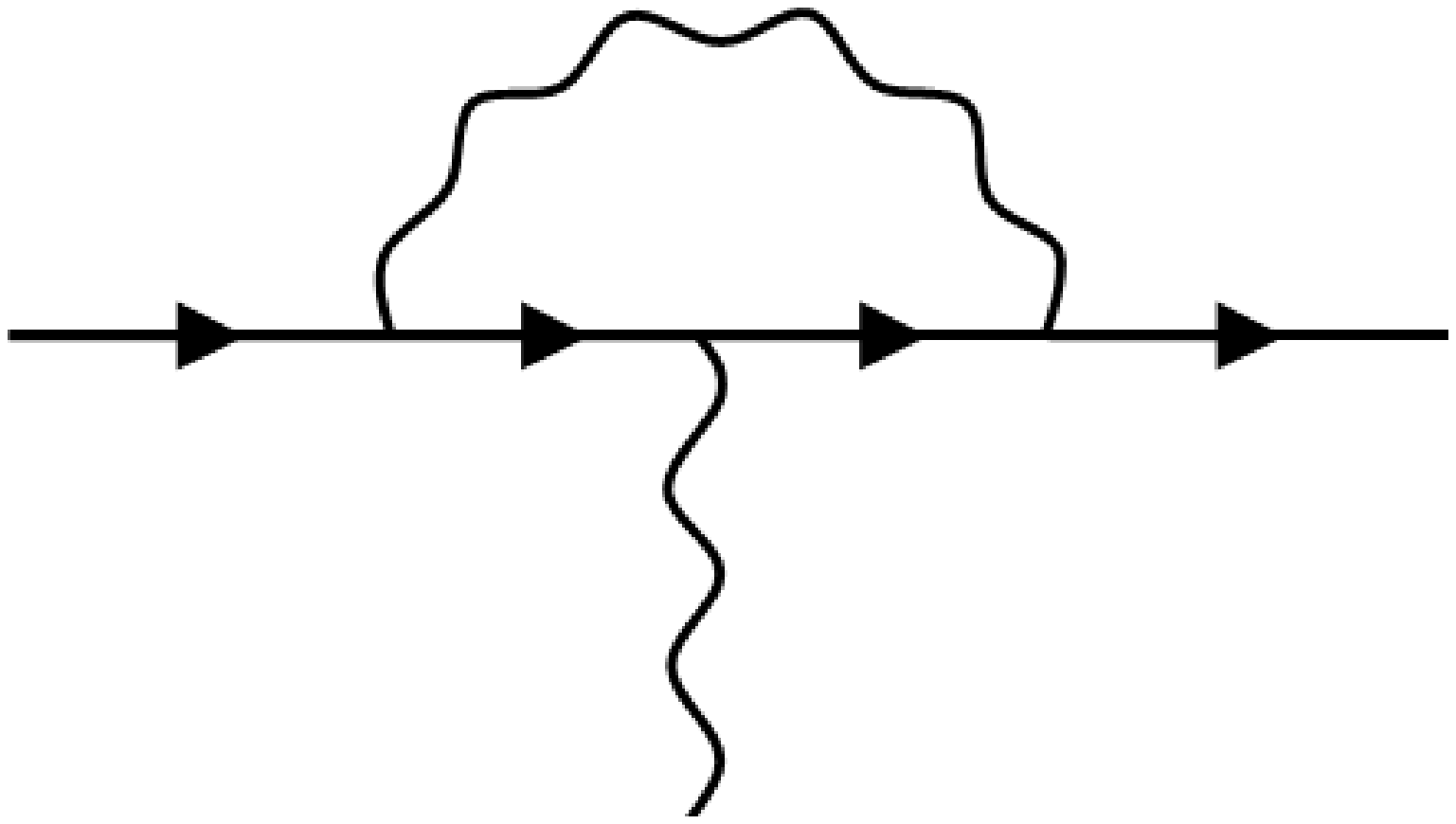} 
\caption{Diagrammatic representation of self energy (a) 
% (Eq.(\ref{eq: self energy})) 
and vertex correction (b) [Eqs.(\ref{eq: self energy}),(\ref{eq: vertex})]. Straight lines with arrows represent fermion propagator, Eq.(\ref{eq: bare Green function}), wavy lines represent dynamically screened long range interaction, Eq.(\ref{eq: Ueff}).}
\label{fig: self energy}
\end{figure}

At first order in the interaction, the fermion Green function acquires a self energy $\Sigma$, represented diagrammatically (to leading order in the interaction) by Fig.\ref{fig: self energy}(a). A self-consistent expression for the change in the fermion propagator $G$ is
\begin{eqnarray}
\label{eq: renormalized green function}
&& \delta G(\omega, \vec{q}) = G_0(\omega, \vec{q})\Sigma(\omega, \vec{q}) G_0(\omega, \vec{q})
, \label{eq: propagator renormalization}\\
&& G_0(\omega, \vec{q}) = \frac{Z_0}{i\omega - H_0(\vec{q})}
,
%LL Z_0 \left(\frac{i\omega + H_0(\vec{q})}{\omega^2 + \big(\frac{q^2}{2m}\big)^2}\right) 
\label{eq: bare Green function}\\
&& \Sigma(\omega, \vec{q}) \! = \! -\!\! \int \!\!\frac{d \epsilon d^2p}{(2\pi)^3} 
% \frac{2 \pi \Gamma_0^2 e^2 G_0(\epsilon + \omega, \vec{p} + \vec{q}) }{\kappa p - 2\pi e^2\Pi(2m \epsilon/p^2)} \label{eq: self energy} .
\Gamma_0^2U_{\epsilon,\vec p}G_0(\epsilon + \omega, \vec{p} + \vec{q})
, \label{eq: self energy}
\end{eqnarray}
%
%Now, the non-interacting Green function is $G_0(\omega, \vec{q}) = (i\omega - H_0(\vec{q}))^{-1}$, which is a $2\times2$ matrix in sublattice space.
%
 %We calculate the interaction renormalization of this Green function by working perturbatively in the interactions, and resumming all one-particle irreducible graphs. This provides a self energy correction to the Green function,
%\begin{equation}G^{-1} (\omega, \mathbf{q}) = G_0^{-1}(\omega, \mathbf{q}) -\Sigma (\omega, \mathbf{q})\label{eq: Green function},\end{equation} where the self energy $\Sigma(\omega, \mathbf{q})$ is represented by Fig.\ref{fig: self energy}, and may be explicitly calculated as 
%
where $\Sigma$ is a $2\times2$ matrix in sublattice space. 

A number of general properties of the self energy can be established based on symmetry considerations. It follows from
% Consideration of 
Eq.(\ref{eq: self energy}) that $\Sigma(0,0)$ vanishes, 
since the part of $G(\epsilon,\vec p)$ which is invariant under rotations of $\vec p$ is an odd function of frequency $\epsilon$. 
Likewise, the expressions for diagonal entries $\Sigma_{AA}(0,\vec{q})$ and $\Sigma_{BB}(0,\vec{q})$, which involve an integral of an odd function of $\epsilon$,
% are odd under $\epsilon \rightarrow - \epsilon$, and 
vanish on integration over $\epsilon$. 
% Meanwhile, 
For the same reason, the expressions for off diagonal entries $\Sigma_{AB}(\omega,0)$ and $\Sigma_{BA}(\omega,0)$ vanish upon integrating the momentum $\vec{p}$ over angles. 
%
% Non trivial effects 
Hence, nonvanishing contributions arise at lowest order when 
% we expand 
the right hand side of Eq.(\ref{eq: self energy}) is expanded to leading order in small $\omega$ and $\vec q$. We obtain%$\Sigma = \omega \partial \Sigma /\partial \omega + \frac{\vec{q}^2}{2m}\partial \Sigma/\partial (\vec{q}^2/2m) + O(\omega^2, q^4, \omega q^2)$. It may be verified that $\partial \Sigma/\partial \omega \propto 1$ and $\partial^2 \Sigma/\partial q^2 \propto H_0$. 
\begin{eqnarray}
%\Sigma(\omega, \vec{q}) = \left( \begin{array}{cc}-i\omega \big(\frac{i\partial \Sigma_{AA}}{\partial \omega}\big) &\frac{\vec{q}_+^2}{2m} \frac{\partial \Sigma_{AB}}{\partial (q_+^2/2m)} \\  \frac{\vec{q}_-^2}{2m} \frac{\partial \Sigma_{BA}}{\partial (q_-^2/2m)}&-i\omega \big(\frac{i\partial \Sigma_{BB}}{\partial \omega}\big) \end{array}\right) + O(\omega^2, q^4) 
\Sigma_{AA}(\omega, \vec{q}) &=& -i \omega \frac{i \partial \Sigma_{AA}(0,0)}{\partial \omega}  + O(\omega^2, \omega q^2, q^4)
, \label{eq: sigmaaa} \\
\Sigma_{AB}(\omega, \vec{q}) &=& \frac{\vec{q}_+^2}{2m} \frac{\partial \Sigma_{AB}(0,0)}{\partial (q_+^2/2m)} + O(\omega^2, \omega q^2, q^4)
, \label{eq: sigmaab}
\end{eqnarray}
where $\Sigma_{AA} = \Sigma_{BB}$ and $\Sigma_{AB} = \Sigma_{BA}^*$ by symmetry.  

It was shown in Ref.[\onlinecite{Yang}] that $i\partial \Sigma_{AA}/\partial \omega$ and $\partial \Sigma_{AB}/\partial (q_+^2/2m)$ are both $\log^2$ divergent, and are equal to leading order (see below and Sec.\ref{sec: mass} for alternative derivation).
% (an explicit calculation is presented below).
% see \cite{supplement}).
Thus the self energy can be written, with log$^2$ accuracy, as 
\begin{equation}
\Sigma(\omega, \vec{q}) = - i Z_0 \frac{\partial \Sigma}{\partial \omega} G_0^{-1}(\omega, \vec{q}) + O\lp \ln \frac{\Lambda}{\Lambda'}\rp
. \label{eq: approx sigma}
\end{equation}
Here, it is understood that non-vanishing $\partial \Sigma/\partial \omega$ is due to the modes that have been integrated out, Eq.(\ref{eq: shell}). Within the leading log approximation, the electron Green function, Eq.(\ref{eq: bare Green function}), retains its non-interacting form, 
% and the self energy Eq.\ref{eq: approx sigma}, 
whereby the self energy, upon substitution into Eq.(\ref{eq: propagator renormalization}),  can be absorbed entirely into a redefinition of the quasiparticle residue, as 
\begin{equation}
% \label{eq: redefinition}% 
\label{eq: renormalization}
\delta G (\omega, \mathbf{q})= \frac{1}{i\omega - H_0(\vec{q})} \delta Z 
%LL \frac{i\omega + H_0(\vec{q})}{\omega^2 + (\frac{q^2}{2m})^2} \delta Z 
, \qquad \delta Z = - i \frac{\partial \Sigma_{AA}}{\partial \omega}Z_0^2.
%\frac{m^*}{m}= \frac{1+i\frac{d\Sigma(0,0)}{d\omega}}{1+\frac{d\Sigma(0,0)}{d(q^2/2m)}}\nonumber 
\end{equation}
We emphasize that the lack of renormalization of the mass only holds at $\log^2$ order. The subleading single log renormalization of the mass will be analyzed in Sec.\ref{sec: mass}. 

% We now determine t
The renormalization of the quasiparticle residue, Eq.(\ref{eq: renormalization}), can be evaluated explicitly by calculating $i\partial \Sigma/\partial \omega$. Taking $\Sigma$ from Eq.(\ref{eq: self energy}), we write
\begin{equation}
i \frac{\partial \Sigma}{\partial \omega} \bigg|_{\omega = 0}= - \int \frac{d\epsilon d^2p}{(2\pi)^3} \frac{(\frac{p^2}{2m})^2 - \epsilon^2}{((\frac{p^2}{2m})^2 + \epsilon^2)^2} \frac{2 \pi \Gamma_0^2 Z_0 e^2}{\kappa p - 2\pi e^2 \Pi(\frac{2m\epsilon}{p^2})}
. \label{eq: res renormalization}
\end{equation}
We express the momenta in polar coordinates $p_x = p \cos \alpha$, $p_y = p \sin \alpha$, and straightaway integrate over $-\pi<\alpha<\pi$. We further change to 
pseudopolar coordinates in the frequency-momentum space, $\epsilon = r\cos \theta$, $p^2/2m = r \sin \theta$, 
with the ``polar angle'' $0<\theta<\pi$.
Using the Rydberg energy  $E_0$, Eq.(\ref{eq: scales}), as units for $r$, we have
% , and measure $r$ in units of $E_0$ (Eq.\ref{eq: scales}), to obtain
%
\begin{equation}
i \frac{\partial \Sigma}{\partial \omega} = - \int^{\Lambda}_{\Lambda'}\frac{dr}{r} \int_{0}^{\pi} \frac{d\theta}{2\pi} \frac{(\sin^2\theta - \cos^2 \theta)\Gamma_0^2 Z_0}{\sqrt{2 r \sin{\theta}} - \frac{2\pi}{m} \Pi(\theta)}, \label{eq: intermediate}
\end{equation}
where $\Pi(\theta)$ is the dimensionless polarization function, given by Eq.(\ref{eq: polfn}) with quasiparticle mass $m$ suppressed and $2m \epsilon/p^2 = \cot \theta$. We note that $\Pi(\theta)$ goes to zero when $\theta \rightarrow 0, \pi$, and these zeros of the polarization function dominate the integral 
% (and provide 
and lead to the $\log^2$ divergence. 
% We also note that 
Since $\Pi(\theta)$ is even about $\theta = \pi/2$, the $\log^2$ contribution can be evaluated by
% . We therefore replace 
replacing $\Pi(\theta)$ in Eq.(\ref{eq: intermediate}) by its asymptotic $\theta\ll \pi$ form,
% as $\theta \rightarrow 0$, 
%
\begin{equation}
\Pi(\theta) \approx \frac{Nm}{4} \tan \theta.
\end{equation}
In the region $\theta \ll \pi$, we may approximate $\sin \theta \approx \theta$, $\tan \theta \approx \theta$ and $\cos \theta \approx 1$. Including a factor of $2$ for the region $\theta \approx \pi$, which gives a contribution identical to that of the region $\theta\approx 0$, we can express the integral Eq.(\ref{eq: intermediate}) with logarithmic accuracy as 
\begin{equation}
i \frac{\partial \Sigma}{\partial \omega} = 2 \int^{\Lambda}_{\Lambda'}\frac{dr}{r} \int_{0}^{\pi/2} \frac{d\theta}{2\pi} \frac{\Gamma_0^2 Z_0}{\sqrt{2 r \theta} + \frac{N \pi}{2} \theta}
.\label{eq: intermediate2}
\end{equation}
Performing the integral over $\theta$ and assuming $r \ll N^2 $ yields
\begin{equation}
i \frac{\partial \Sigma}{\partial \omega} = \frac{2\Gamma_0^2Z_0}{N\pi^2} \int_{\Lambda'}^{\Lambda} \frac{dr}{r} \ln \frac{N^2 \pi^2}{8 r}.
\end{equation}
Integrating over $\Lambda'<r<\Lambda$ 
% the shell 
(see Eq.(\ref{eq: shell})), we obtain
\begin{equation}
i \frac{\partial \Sigma}{\partial \omega} = \frac{2 \Gamma_0^2Z_0}{N \pi^2} \lp \ln \frac{N^2 \pi^2 E_0}{8 \Lambda'}\ln \frac{\Lambda}{\Lambda'}  - \frac{1}{2}\ln^2 \frac{\Lambda}{\Lambda'}\rp \label{eq: Z}
. \end{equation}
%
%\mpar{$\Lambda_1 = \Lambda_0 - \delta \Lambda$ suppressed}
We now consider an infinitesimal RG transformation.
%LL $\Lambda_1 = \Lambda_0 - \delta \Lambda$. %In this limit, we obtain
%
%\begin{equation}
%Z_1 = Z_0 - \frac{2 \Gamma_0^2 Z_0^3}{N\pi^2} \ln \frac{N^2 E_0}{\Lambda_0} \ln \frac{\Lambda_0}{\Lambda_1} \label{eq: Z} \end{equation}
%
% We identify 
Defining an RG time 
%
%\mpar{OK?}
\begin{equation}
%LL \xi = \ln \frac{N^2 \pi^2 E_0}{8 \Lambda_0}
\xi = \ln \frac{\Lambda_0}{\Lambda}
, \qquad \delta \xi = \ln \frac{\Lambda}{\Lambda'} \label{eq: RG time},
\end{equation}
% and hence 
we rewrite the recursion relation, Eq.(\ref{eq: Z}), as
%
%\mpar{OK?}
\begin{equation}
i \frac{\partial \Sigma}{\partial \omega} = \frac{2 \Gamma_0^2Z_0}{N \pi^2} (\xi +c) d\xi 
,\quad
c=\ln \frac{N^2 \pi^2 E_0}{8 \Lambda_0}
.
\label{eq: difference equation}
\end{equation}
The constant term $c$ describes corrections subleading in $\log^2$, and thus may seem to be irrelevant. However, we shall retain it in the RG equation since it will determine the form of renormalization near the UV cutoff (see discussion of TDOS in Sec.\ref{sec: dos}). 

In our derivation of Eq.(\ref{eq: difference equation}) it was assumed that our initial UV cutoff $\Lambda_0 < N^2 \pi^2 E_0/8$. Such choice of $\Lambda_0$ is certainly justified when $N$ is large, which is the limit we worked in thus far. Better still, the condition remains entirely reasonable for the physical value $N=4$, leading to $N^2 \pi^2 E_0/8 = 24 eV \kappa^{-2}$, which is much bigger than the bandwidth for BLG. %\mpar{OK?}

Substituting Eq.(\ref{eq: difference equation}) into Eq.(\ref{eq: renormalization}), we obtain a differential equation for the flow of the quasiparticle residue, 
\begin{equation}
\frac{\partial Z}{\partial \xi} = - \frac{2 \Gamma^2(\xi) Z^3(\xi)}{N\pi^2} (\xi +c)
. \label{eq: z}
\end{equation}
This equation encapsulates a one loop RG flow for the residue $Z$, describing its renormalization within 
% logarithmic
% leading 
a $\log^2$ accuracy. 

\section{Self-consistent renormalization of the vertex function $\Gamma$}
\label{sec5}

The screened Coulomb interaction renormalizes through the vertex correction, pictured in Fig.\ref{fig: self energy}(b). The RPA 
% loop has 
bubble diagrams, which have already been taken into account in 
% the screened interaction, Eq.(\ref{eq: Ueff}),
moving from an unscreened to a screened interaction, Eq.(\ref{eq: Ueff}),
% and so does 
do not contribute to renormalization.
It may be verified by an explicit calculation that the vertex correction in Fig.\ref{fig: self energy}(b) is given by 
\begin{equation}
\delta \Gamma = - \int \frac{d\epsilon d^2p}{(2\pi)^3} \frac{(\frac{p^2}{2m})^2 - \epsilon^2}{((\frac{p^2}{2m})^2 + \epsilon^2)^2} \frac{2 \pi \Gamma_0^3 Z_0^2 e^2}{\kappa p - 2\pi e^2 \Pi(\frac{2m\epsilon}{p^2})}. \label{eq: vertex}
\end{equation}
This is the same expression as for the residue renormalization [Eqs.(\ref{eq: renormalization}),(\ref{eq: res renormalization})], 
with $\Gamma$ replacing $Z$, and a sign change.
% with the opposite sign. 
Hence, we obtain
\begin{equation}
\frac{\partial \Gamma}{\partial \xi} = \frac{2\Gamma^3(\xi) Z^2(\xi)}{N\pi^2} (\xi+c) \label{eq: Gamma}
\end{equation}
which is identical to the flow equation for $Z$, albeit with a reversed sign.
% , with substitution $Z \leftrightarrow \Gamma$, and the opposite sign. 
Therefore, the product $\Gamma Z$ does not renormalize at log square order, and we can write. 
\begin{equation}
\Gamma(\xi) Z(\xi) = 1 \label{eq: Ward identity}. 
\end{equation}
This result is not a coincidence, since the residue $Z$ and the vertex function $\Gamma$ are not independent quantities. The Hamiltonian, Eq.(\ref{eq: Hamiltonian}), is invariant under a gauge transformation
% global phase shift of 
of electron wavefunction $\psi'=\psi e^{i\chi}$, accompanied by energy and momentum shifts $\epsilon'=\epsilon-\partial_t\chi$, $\vec p'=\vec p+\nabla\chi$.
% rotation $\psi \rightarrow \psi e^{i\theta}$. 
%\mpar{drop particle number conservation; this is really just the old good gauge invariance}
This gauge invariance symmetry can be shown to lead to Eq.(\ref{eq: Ward identity}) through a Ward identity that relates the self-energy to the vertex function [\onlinecite{Landau, Gonzalez}].  %The action Eq.\ref{eq: partition function} is invariant under the electromagnetic gauge transformations 
%
%\begin{equation}
%\psi \rightarrow \psi \exp(i \theta) \qquad \phi \rightarrow \phi - \frac{\partial \theta/\partial t}{e \Gamma Z} \label{eq: gauge transformations}.
%\end{equation}
%
%Here, the gauge transformation $\theta$ is spatially uniform. This is necessary, since according to the action Eq.\ref{eq: coupling},\ref{eq: dressed propagator}, spatial gradients in the field $\phi$ correspond to (screened) electric fields, which are physical. However, time derivatives of $\phi$ have no physical meaning, and represent a gauge symmetry. 
%This gauge invariance demands \cite{Gonzalez} that $\Gamma$ and $Z$ are related (at leading log square order) by the Ward identity 

% STOPPED HERE

\section{Renormalization of the four point vertex $\Upsilon$}
\label{sec6}

\begin{figure}
\large{a)} \includegraphics[scale = 0.26]{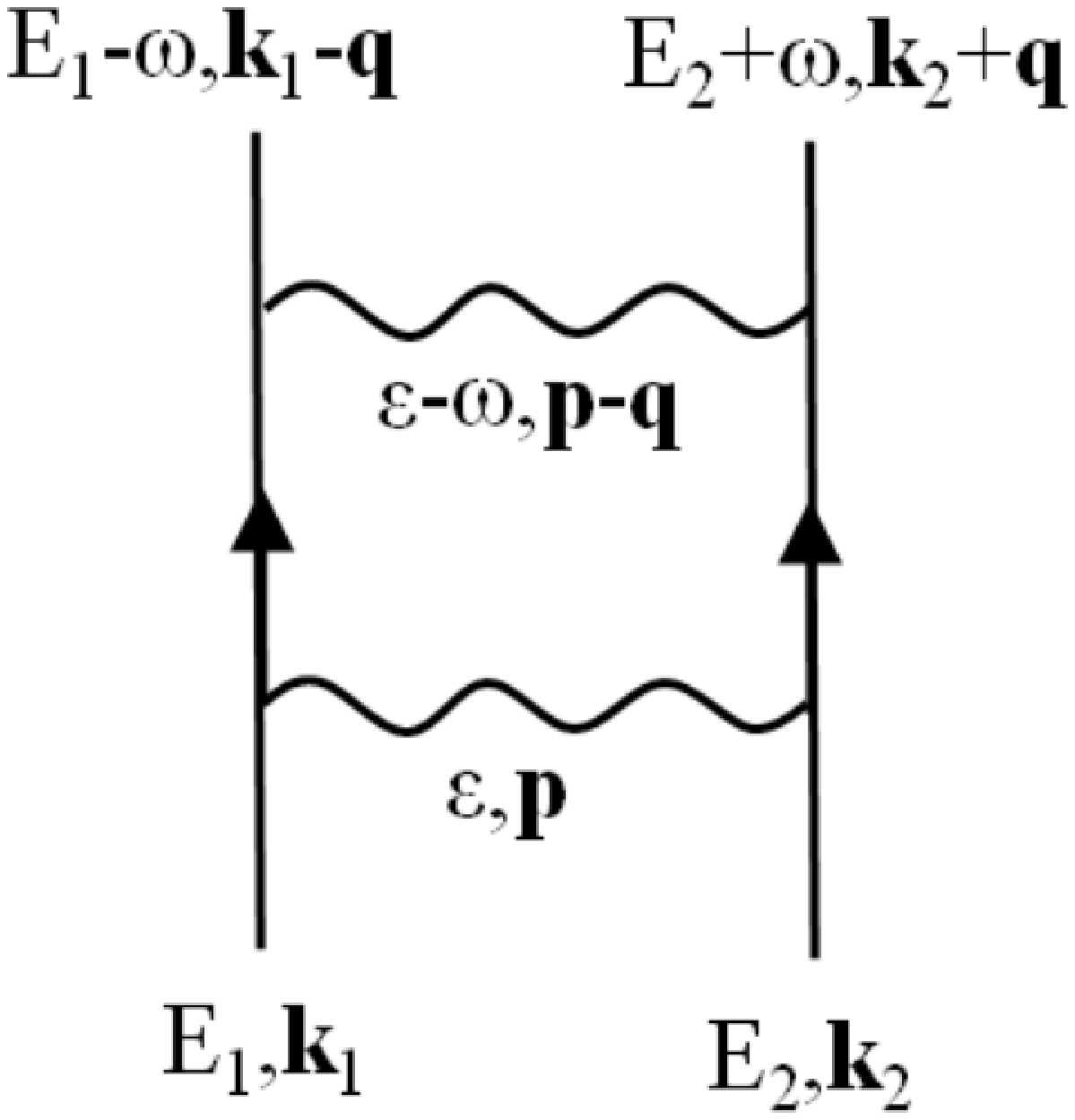}
\large{b)} \includegraphics[scale = 0.56]{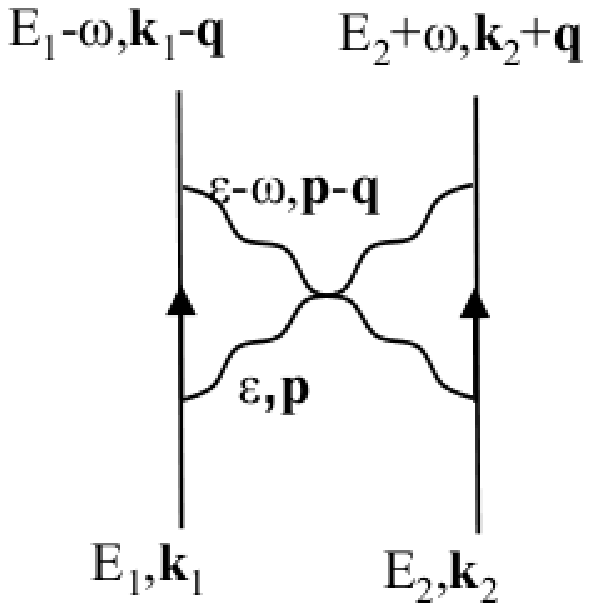}
\caption{The renormalization of the four point vertex $\Upsilon$ proceeds through repeated scattering in the particle particle channel (a) and in the particle hole channel (b), known as the the BCS loop and the ZS' loop in the Luttinger liquid literature \cite{Shankar}.  
%. We use the same nomenclature as the literature on Luttinger liquids \cite{Shankar} and call these processes the ZS' loop (a) and BCS loop (b) respectively. 
The RPA bubble diagrams (ZS loop in the language of Ref.[\onlinecite{Shankar}]), which arise in the same order of perturbation theory, have already been taken into account in the screened interaction, Eq.(\ref{eq: Ueff}).}
% when screening the Coulomb interaction, 
% and should not be counted a second time.}
\label{fig: interaction renormalization}
\end{figure}

The four point vertex $\Upsilon$, introduced in Eq.(\ref{eq: four point vertex}), renormalizes through the diagrams presented in Fig.\ref{fig: interaction renormalization}(a,b), which represent the repeated scattering of two particles in the electron-electron and electron-hole channels respectively.  We follow the naming conventions 
% introduced 
used in Ref.[\onlinecite{Shankar}] in the context of the Luttinger liquid, and name these two diagrams, the BCS loop and the ZS' loop, pictured in Fig.\ref{fig: interaction renormalization}(a) and Fig.\ref{fig: interaction renormalization}(b), respectively. In the one dimensional Luttinger liquids, the two processes famously cancel\cite{DzyaloshinskiiLarkin}, so that the four point vertex does not renormalize. In higher dimensions, such a cancellation is rare. However, the discrete nature of the Fermi surface in BLG 
% opens the door to 
results in a Luttinger liquid like cancellation of the processes Fig.\ref{fig: interaction renormalization}(a,b), as will be discussed below. 

We argued in Sec.\ref{sec: setup} that the RG-relevant scattering processes allowed by symmetry must have sublattice structure $(A,A)\rightarrow (A,A)$, $(A,B) \rightarrow (A,B)$ or $(A,B) \rightarrow (B,A)$. 
To see the mathematical origin of such selection,
%LL At this point, 
it is instructive to explicitly write out the form of the electron Green function. We have 
\begin{eqnarray}
&& G_{AA}(\epsilon, \vec{p}) = \frac{-Z i\epsilon}{\epsilon^2 + \lp p^2/2m\rp^2} = G_{BB}(\epsilon, \vec{p}), \label{eq: gaa} \\
&& G_{AB}(\epsilon, \vec{p}) = \frac{-Z p_+^2/2m}{\epsilon^2 + \lp p^2/2m\rp^2} = G^*_{BA}(\epsilon, \vec{p}).\label{eq: gab}
\end{eqnarray}
 When the diagrams Fig.\ref{fig: interaction renormalization}(a,b) are evaluated in any channel other than these three channels,
%LL  symmetry allowed, 
they vanish upon integration over inner momentum variables, due to the chiral structure of the sublattice changing Green functions, Eq.(\ref{eq: gab}). 
%LL \mpar{OK?} 
%causes the integrals represented by the diagrams in Fig.\ref{fig: interaction renormalization} to vanish identically, \mpar{vanish identically? is that clear?}
%unless the scattering process involved is `sublattice conserving' \cite{Zhang}. That is, the only contribution is to $\Upsilon_{iijj}$ and $\Upsilon_{ijji}$, where $i$ and $j$ are sublattice indices. This result was already anticipated on symmetry grounds in the discussion surrounding Eq.(\ref{eq: four point vertex}). 

Similar reasoning leads to a conclusion 
%LL We now show 
that the $(A,B) \rightarrow (B,A)$ vertex cannot exhibit a  $\log^2$ divergence. 
As we saw above,
%LL channel does not contribute at leading $\log^2$ order. 
the $\log^2$ divergences arise because the effective interaction $U_{\epsilon, \vec{p}}$ has a pole at $\vec p=0$ and finite $\epsilon$. However, the sublattice changing Green functions, Eq.(\ref{eq: gab}), have zeros at small $\vec p$, which cancel the contribution of the pole in the interaction. Thus, the diagrams in Fig.\ref{fig: interaction renormalization} can only be $\log^2$ divergent if all internal Green functions are sublattice preserving, given by Eq.(\ref{eq: gaa}). Since the process $(AB) \rightarrow (BA)$ involves two sublattice changing Green functions, it follows that the integrals associated with this processes cannot be $\log^2$ divergent, and hence this process does not contribute at leading $\log^2$ order.
%\mpar{?}
%This means that the diagrams in Fig.\ref{fig: interaction renormalization} cannot have any contribution i%n the exchange channel $\Upsilon_{ABBA}$ at leading order ($\log^2$). 
%\mpar{Does this supercede an earlier discussion above and in Sec.III? what's the relation between them?}
%We may therefore concentrate on the contribution to the direct channel $\Upsilon_{iijj}$. 

Thus, at leading order, we need to consider only the processes $(AA) \rightarrow (AA)$ and $(AB) \rightarrow (AB)$. Moreover, since the interaction (\ref{eq: Ueff}) does not distinguish between sublattices, the ZS' and BCS contributions from Fig.\ref{fig: interaction renormalization}(a,b) in these
%LL the $(AA) \rightarrow (AA)$ and $(AB) \rightarrow (AB)$ 
channels are the same. Therefore, to demonstrate that $\Upsilon$ does not renormalize at leading order, it is sufficient to demonstrate that there are no $\log^2$ divergences in the $(AA) \rightarrow (AA)$ channel. 
%In addition, since the interactions do not distinguish between sublattices, and $G_{AA} = G_{BB}$, it also follows that the contribution to $\Upsilon_{iijj}$ from the diagrams in Fig.\ref{fig: interaction renormalization} does not depend on sublattice type. We may therefore restrict ourselves to evaluating the contribution of the ZS' and BCS diagrams (Fig.\ref{fig: interaction renormalization}) in the specific instance where the incoming and outgoing particles are all on sublattice A. The flavour of the interacting particles is immaterial, due to the SU(4) flavour symmetry of the Hamiltonian Eq.(\ref{eq: Hamiltonian}).

In evaluating the ZS' and BCS diagrams (Fig.\ref{fig: interaction renormalization}), it will prove important to keep track of external momenta. The vertex $\Upsilon(E_1, E_2, \omega, \vec{k}_1, \vec{k}_2, \vec{q})$ then represents the amplitude for the scattering process 
\[
\psi_{\sigma, A, E_1, \vec{k}_1} \psi_{\sigma', A, E_2, \vec{k}_2} \rightarrow \psi_{\sigma, A, E_1+ \omega, \vec{k}_1 + \vec{q}} \psi_{\sigma', A, E_2 - \omega, \vec{k}_2 - \vec{q}}.
\] 
Translating the ZS' and BCS diagrams in Fig.\ref{fig: interaction renormalization} into integrals, we find the contributions 
%LL to the vertex $\Upsilon(\omega, \omega', \vec{q}, \vec{q'})$ from the ZS' and BCS diagrams is 
%
%\begin{widetext}
\begin{eqnarray}
\Upsilon_{AAAA}^{\rm{ZS'}} &=& \Gamma^4 \int \frac{d\epsilon d^2p}{(2\pi)^3} U_{\epsilon, \vec{p}} U_{\epsilon - \omega, \vec{p} - \vec{q}} 
% G_{AA}(\epsilon_1,\vec p_1)G_{AA}(\epsilon_2,\vec p_2)
G_{AA}(E_1 + \epsilon, \vec{k}_1 + \vec{p})
\nonumber \\
&&\times G_{AA}(E_2 + \epsilon - \omega, \vec{k}_2 + \vec{p} - \vec{q})
,
\label{eq: zs'}
\\
% \end{eqnarray}
% \begin{eqnarray}
\Upsilon_{AAAA}^{\rm{BCS}} &=& \Gamma^4 \int \frac{d\epsilon d^2p}{(2\pi)^3} U_{\epsilon, \vec{p}} U_{\epsilon - \omega, \vec{p} - \vec{q}} 
% G_{AA}(\epsilon_1,\vec p_1)G_{AA}(\tilde \epsilon_2,\tilde {\vec p}_2)
G_{AA}(E_1 + \epsilon, \vec{k}_1 + \vec{p})\nonumber\\
&& \times G_{AA}(E_2 - \epsilon, \vec{k}_2 - \vec{p}) 
.
\label{eq: BCS}
\end{eqnarray}
%\end{widetext}
%
%We have introduced the notation $\epsilon_{\pm} = \epsilon \pm \omega$ and $\vec{p}_{\pm} = \vec{p} \pm \vec{q}$. 
Here, the interaction $U(\epsilon, p)$ is defined by Eq.(\ref{eq: Ueff}), the Green functions are defined by Eq.(\ref{eq: gaa}), and the integral goes over the shell defined by Eq.(\ref{eq: shell}). 

As always in a RG analysis, we assume that the external frequencies and momenta are small compared to the internal frequencies and momenta:
\begin{equation}\label{eq: condition12}
\max\lp \omega, \omega', \frac{{\vec q}^2}{2m}, \frac{{\vec q}'^2}{2m}\rp \ll \Lambda' < \sqrt{\epsilon^2 + \lp \frac{{\vec p}^2}{2m}\rp ^2} <\Lambda
.
% &\ge& \Lambda_1 \label{eq: condition1}\\
% \max(\omega, \omega', \frac{q^2}{2m}, \frac{q'^2}{2m}) &\ll& \Lambda_1 \label{eq: condition2}
\end{equation}
In such a case, the standard approach to handling the integrals over $\epsilon$ and $\vec p$ involves setting the external frequency and momenta to zero at first, and restoring their finite values later to regulate the infrared (IR) divergences.
However, a straightforward application of this recipe to the integrals in Eqs.(\ref{eq: zs'}),(\ref{eq: BCS}) proves impossible, 
%LL At this point there is cause for concern, 
because these integrals are power law divergent when all external momenta are set to zero.
%LL , then the expressions in Eqs.(\ref{eq: zs'}),(\ref{eq: BCS}) are power law divergent. This strong 
The divergence arises from the region near $\vec p \approx 0$ (which lies within the shell defined by Eq.(\ref{eq: shell})), where the interaction is nearly unscreened. In this region, we have
%\mpar{equation OK?}
%
\begin{equation}
U_{\epsilon, \vec{p}} U_{\epsilon - \omega, \vec{p} - \vec{q}} \sim 
\frac{1}{\lp |\vec p| + \alpha |\vec p|^2 \rp \lp |\vec{p}-\vec{q}| + \alpha |\vec{p}-\vec{q}|^2\rp}
,
%LL \frac{1}{\lp \frac{\kappa |\vec p|}{e^2} + \frac{N |\vec p|^2}{2\Lambda_0}\rp\lp\frac{\kappa |\vec{p}-\vec{q}|}{e^2} + \frac{N|\vec{p}-\vec{q}|^2}{2\Lambda_0}\rp}
\end{equation}
with $\alpha= Ne^2/2\kappa\Lambda$. At finite $\vec{q}$, the poles in this expression are split apart, and thus the singular contribution of each pole, $\vec p=0$ and $\vec p=\vec q$,
% line has a pole, but this pole 
is regularized by the integration measure $d^2p$ so that the integrals in Eqs.(\ref{eq: zs'}),(\ref{eq: BCS}) remain well defined. However, when all external momenta are zero, the poles from the two interaction lines co-incide, and the expressions (\ref{eq: zs'}), (\ref{eq: BCS}) acquire a second order pole at $\vec p=0$. When we integrate over this second order pole, we pick up a power law divergence. 

Hence, if either of the ZS' or BCS diagrams existed in isolation, this power law divergence would indicate a strong (power law) instability, which would drive $\Upsilon$ into the strong coupling regime, where our $\log^2$ RG would cease to apply. However, as we will now show, the divergences in the contributions to $\Upsilon$ from the expressions (\ref{eq: zs'}), (\ref{eq: BCS}) in fact cancel out, so that $\Upsilon$ does not flow to $\log^2$ order. %\mpar{repeat?}
%This cancellation has strong parallels to the cancellation between particle-particle and particle-hole scattering in the one dimensional Luttinger liquid \cite{Shankar}.
%LL We wish to confirm by careful calculation that the contributions to the four point vertex from scattering in the particle-particle and particle hole channels do indeed cancel at $\log^2$ order. Now t
To analyse the cancellation between the ZS' and BCS diagrams, it is convenient to add the integrands of Eq.(\ref{eq: zs'}) and Eq.(\ref{eq: BCS}) together before doing the integral, while keeping external momenta finite. Preserving finite external momenta ensures that the integrals Eq.(\ref{eq: zs'}) and Eq.(\ref{eq: BCS}) are well defined. After combining the integrands, and denoting $\Upsilon_{AAAA}^{ZS'}+\Upsilon_{AAAA}^{BCS}=\tilde\Upsilon$, we obtain
%LL where $\tilde\Upsilon$ denotes the integral

\begin{widetext}
\begin{equation}
% \Upsilon_{AAAA}^{ZS' + BCS} 
\tilde\Upsilon = \Gamma^4 \int \frac{d\epsilon d^2p}{(2\pi)^3} U_{\epsilon, \vec{p}} U_{\epsilon - \omega, \vec{p} - \vec{q}} G_{AA}(E_1 + \epsilon, \vec{k_1} + \vec{p}) \left[G_{AA}(E_2 + \epsilon - \omega, \vec{k}_2 + \vec{p} - \vec{q}) + G_{AA}(E_2 - \epsilon, \vec{k}_2 - \vec{p}) \right]
. \label{eq: added}
%\frac{Z^2 i(\omega' - \epsilon)}{(\omega' - \epsilon)^2 + \lp \frac{|\vec{q'} - \vec{p}|^2}{2m}\rp ^2} \left(\frac{i\epsilon}{\epsilon^2 + \lp \frac{p^2}{2m}\rp ^2} + \frac{i(\omega - \epsilon)}{(\omega - \epsilon)^2 + \lp \frac{|\vec{q} - \vec{p}|^2}{2m}\rp ^2}\right) \label{eq: added}
\end{equation}
\end{widetext}
%
%LL Now, we further  simplify by noting
To simplify this expression we note that momentum $\vec q$ enters very differently in Eq.(\ref{eq: added}) as compared to other external frequencies and momenta $E_1$, $E_2$, $\omega$ , $\vec{k}_1$, $\vec{k}_2$.
% , $(E_1, E_2, \omega, \vec{k}_1, \vec{k}_2)$ enter very differently in Eq.(\ref{eq: added}) to $\vec{q}$. 
The momentum $\vec{q}$ is needed to split the poles coming from the two interaction terms -- if we take $\vec{q}$ to zero, the integral will acquire a second order pole at $\vec p = 0$, leading to a divergence. This divergence arises from within the shell that we are integrating out (Eq.(\ref{eq: shell})), and thus the RG will be ill defined. In contrast, sending the frequencies and momenta $E_1$, $E_2$, $\omega$, $\vec{k}_1$, $\vec{k}_2$ to zero by applying Eq.(\ref{eq: condition12}) does not cause any concern. We thus have
%LL , can be safely sent to zero 
% ,(\ref{eq: condition2}), 
%LL to obtain
%
% \begin{widetext}
\begin{eqnarray}
% \Upsilon_{AAAA}^{ZS' + BCS} 
\tilde\Upsilon &=& \Gamma^4 \int \frac{d\epsilon d^2p}{(2\pi)^3} U_{\epsilon, \vec{p}} U_{\epsilon, \vec{p} - \vec{q}} G_{AA}(\epsilon,\vec{p}) \nonumber\\
&&\times \left[G_{AA}(\epsilon, \vec{p} - \vec{q}) + G_{AA}( - \epsilon, - \vec{p}) \right]. \label{eq: target}
%\frac{Z^2 i(\omega' - \epsilon)}{(\omega' - \epsilon)^2 + \lp \frac{|\vec{q'} - \vec{p}|^2}{2m}\rp ^2} \left(\frac{i\epsilon}{\epsilon^2 + \lp \frac{p^2}{2m}\rp ^2} + \frac{i(\omega - \epsilon)}{(\omega - \epsilon)^2 + \lp \frac{|\vec{q} - \vec{p}|^2}{2m}\rp ^2}\right) \label{eq: added}
\end{eqnarray}
% \end{widetext}
%
Interestingly, the expression in square brackets vanishes identically when $\vec{q} = 0$, since $G_{AA}(\epsilon, \vec{p}) = - G_{AA}(-\epsilon, -\vec{p})$. However, taking the limit $\vec{q} \rightarrow 0$ is potentially problematic because of the pole structure of $U_{\epsilon, \vec{p}} U_{\epsilon - \omega, \vec{p} - \vec{q}}$ discussed above.
%LL since in this limit the integrals in Eqs.(\ref{eq: zs'}),(\ref{eq: BCS}) cease to be well defined. 
%The danger stems from the region around $p^2 \approx 0$, (which lies inside the shell Eq.\ref{eq: shell}), where the two interaction lines have poles. 
%LL To demonstrate that the $\Upsilon$ vertex indeed does not renormalize to $\log^2$ order, we must 
Instead, we proceed with caution, and evaluate Eq.(\ref{eq: target}) at finite $\vec{q}$, using the conditions (\ref{eq: condition12})
%, (\ref{eq: condition2}) 
to simplify the analysis. 

Given what we just said, it is now easy to see why there is no $\log^2$ divergence in $\tilde\Upsilon$.
%LL Eq.\ref{eq: target}. 
First, we note that the interaction (\ref{eq: Ueff}) carries a soft UV cutoff, so the integral in Eq.(\ref{eq: target}) is UV convergent (this property of dynamically screened interaction in BLG is discussed e.g. in Ref.[\onlinecite{Nandkishore}]). Hence, we can shift variables to $\vec{p}_\pm = \vec{p}\pm \vec{q}/2$ and rewrite the expression (\ref{eq: target}) as
%LL for $\tilde\Upsilon$ as
%LL \mpar{possible SDS reference?}
% \begin{widetext}
%% \begin{equation}
% \Upsilon_{AAAA}^{ZS' + BCS} 
%% \tilde\Upsilon = \Gamma^4 \int \frac{d\epsilon d^2p}{(2\pi)^3} U_{\epsilon, \vec{p}_+} U_{\epsilon, \vec{p}_-} G_{AA}(\epsilon,\vec{p}_+) \left[G_{AA}(\epsilon, \vec{p}_- ) + G_{AA}( - \epsilon, - \vec{p}_+) \right] \label{eq: added21}
%\frac{Z^2 i(\omega' - \epsilon)}{(\omega' - \epsilon)^2 + \lp \frac{|\vec{q'} - \vec{p}|^2}{2m}\rp ^2} \left(\frac{i\epsilon}{\epsilon^2 + \lp \frac{p^2}{2m}\rp ^2} + \frac{i(\omega - \epsilon)}{(\omega - \epsilon)^2 + \lp \frac{|\vec{q} - \vec{p}|^2}{2m}\rp ^2}\right) \label{eq: added}
%% \end{equation}
%
%% where $\vec{p}_\pm = \vec{p}\pm \vec{q}/2$. We write 
%% \begin{eqnarray}
%% G_{AA}(\epsilon, p) = iZ \epsilon D(\epsilon, \vec{p}) \qquad D(\epsilon, \vec{p}) = \frac{1}{\epsilon^2 + \lp{\vec p}^2/2m\rp^2}
%% \end{eqnarray}
%% and hence rewrite Eq.(\ref{eq: added21}) as 
\begin{widetext}
\begin{eqnarray}
% \Upsilon_{AAAA}^{ZS' + BCS} 
\tilde\Upsilon &=& - \Gamma^4 Z^2  \int \frac{d\epsilon d^2p}{(2\pi)^3} U_{\epsilon, \vec{p}_+} U_{\epsilon, \vec{p}_-} \epsilon^2 D(\epsilon,\vec{p}_+) \left[D(\epsilon, \vec{p}_- ) - D(\epsilon, \vec{p}_+) \right] \\
&=& - \Gamma^4 Z^2 \int \frac{d\epsilon d^2p}{(2\pi)^3} U_{\epsilon, \vec{p}_+} U_{\epsilon, \vec{p}_-} \epsilon^2 \left[\frac{D(\epsilon, \vec{p}_+) + D(\epsilon, \vec{p}_-)}{2} + \frac{D(\epsilon, \vec{p}_+) - D(\epsilon, \vec{p}_-)}{2}\right] \left[D(\epsilon, \vec{p}_- ) - D(\epsilon, \vec{p}_+) \right]
, \nonumber
\end{eqnarray}\end{widetext}
where we factored the Green functions as
\be
G_{AA}(\epsilon, p) = iZ \epsilon D(\epsilon, \vec{p}), \ \ D(\epsilon, \vec{p}) = \frac{1}{\epsilon^2 + \lp{\vec p}^2/2m\rp^2}.
\ee
We note that because $\Upsilon$ should be even under $\vec{q} \rightarrow - \vec{q}$ the first term in the brackets gives zero upon integration over $\vec p$. Hence, we can rewrite the result for $\tilde \Upsilon$, Eq.(39), as 
\begin{eqnarray}
% \Upsilon_{AAAA} 
\tilde\Upsilon &=&  \frac{\Gamma^4Z^2}{2} \int \frac{d\epsilon d^2p}{(2\pi)^3} U_{\epsilon, \vec{p}_+} U_{\epsilon, \vec{p}_-} \epsilon^2 \left[D(\epsilon, \vec{p}_- ) - D(\epsilon, \vec{p}_+) \right]^2 \nonumber \\
&=& \frac{\Gamma^4Z^2}{2} \int \frac{d\epsilon d^2p}{(2\pi)^3} U_{\epsilon, \vec{p}_+} U_{\epsilon, \vec{p}_-} \epsilon^2 \left[\frac{z_+^2 - z_-^2}{(\epsilon^2 + z_+^2)(\epsilon^2 + z_-^2)}\right]^2 
, \nonumber\\ \label{eq: sqbracket} %\nonumber 
% \qquad \qquad z_{\pm} = \frac{|\vec{p}_{\pm}|^2}{2m}
\end{eqnarray}
where $z_{\pm} = |\vec{p}_{\pm}|^2/2m$.
%We combine the two Green functions in square brackets using Eq.\ref{eq: gaa}. We keep only terms to leading order in $\omega$ and $q^2/2m$, and obtain %Since the integrand Eq.\ref{eq: added2} is even under $\omega \rightarrow - \omega$ and $\vec{q} \rightarrow -\vec{q}$, all terms linear in $\omega$ and $\vec{q}$ vanish, so that we obtain%The term The two terms in square brackets may then be expanded to leading order in small . We repeatedly use the conditions Eq.\ref{eq: condition1} and Eq.\ref{eq: condition2} to simplify the expressions, and keep only terms to leading order in $\omega/\Lambda_1$ and $q^2/2m\Lambda_1$. We use Eq.\ref{eq: Ueff} and \ref{eq: ueff2} to obtain
%

To extract the leading contribution at small $\vec{q}$, we approximate the effective interaction as
% , and to write
\begin{equation}
U(\epsilon, \vec{p}) = -\frac{\Pi^{-1}(\epsilon, \vec{p})}{1 - \frac{\kappa |\vec p|}{2\pi e^2 \Pi(\epsilon, \vec{p})}}\approx - \frac{1}{\Pi(\epsilon, \vec{p})}
.
% -\big(\Pi(\epsilon, \vec{p})\big)^{-1} \left(1 - \frac{\kappa p}{2\pi e^2 \Pi(\epsilon, \vec{p})}\right)^{-1} \approx - \frac{1}{\Pi(\epsilon, \vec{p})} 
\label{eq: Uapprox}
\end{equation}
From the definition of the polarization function, Eq.(\ref{eq: polfn}), we see that the approximation $U \approx -1/\Pi$ holds everywhere in the shell Eq.(\ref{eq: shell}) except at $\vec p \approx 0$, since $\Pi(\vec p=0) = 0$. 
However, in the limit $\vec p \rightarrow 0$, the expression in brackets in Eq.\ref{eq: sqbracket}
%LL integrand of $\tilde\Upsilon$
%LL of Eq.(\ref{eq: almost there}) 
tends to zero because of the expansion $z_+^2-z_-^2=(\vec p^2/m)(\vec p\cdot\vec q/2m)+O(\vec p^4)$, which ensures validity of the approximation (\ref{eq: Uapprox}). 

Hence, using Eq.(\ref{eq: polfn}), we obtain
\begin{eqnarray}
% \Upsilon_{AAAA} 
\tilde\Upsilon 
= &&\Gamma^4Z^2 \int d\epsilon d^2p
%LL U_{\epsilon, \vec{p}_+} U_{\epsilon, \vec{p}_-} 
\frac{ \sqrt{(z_+^2+u\epsilon^2)(z_-^2+u\epsilon^2)}}{4\pi(Nm\ln 4)^2 } \nonumber \\
&& \times \frac{\epsilon^2}{z_+z_-}\left[\frac{z_+^2 - z_-^2}{(\epsilon^2 + z_+^2)(\epsilon^2 + z_-^2)}\right]^2 
.
% \qquad \qquad z_{\pm} = \frac{|\vec{p}_{\pm}|^2}{2m}
\end{eqnarray}
Simple power counting shows that this integral is UV convergent, IR convergent, and is completely independent of $q$, which can be scaled out by defining new variables $p' = p/q$ and $\epsilon' = 2m \epsilon/q^2$. It follows that the the diagrams representing repeated scattering in the particle-particle and particle-hole channels do indeed cancel, so that $\Upsilon_{AAAA}Z^2$ does not renormalize. 

Combining this with our argument demonstrating that $\Upsilon_{ABBA}Z^2$ does not renormalize at $\log^2$ order (see discussion below Eq.(32)), and recalling that $\Upsilon_{AAAA}=\Upsilon_{AABB}$, we conclude that we can set $\Upsilon = 0$ with log$^2$ accuracy. 

\section{Solution of RG flow equations. Zero bias anomaly in bilayer graphene}
\label{sec: dos}

%LL We have shown that 
Since the only quantities which renormalize at $\log^2$ order in a one loop RG are the quasiparticle residue $Z$ and the interaction vertex function $\Gamma$, 
%LL and these are related by the Ward identity Eq.(\ref{eq: Ward identity}), 
the problem of finding the RG flow of these quantities reduces to solving Eqs.(\ref{eq: z}),(\ref{eq: Gamma}). All other quantities do not renormalize at log square order, and may thus be treated as constants with logarithmic accuracy. 
%LL The problem of determining the RG flow of BLG for all $\xi$ thus reduces to the problem of solving Eqs.(\ref{eq: z}),(\ref{eq: Gamma}). 

Additional simplification arises due to the Ward identity $\Gamma Z=1$, Eq.(\ref{eq: Ward identity}). Using it to decouple the RG equations for $Z$ and $\Gamma$, we write the equation for $Z$ as
\be\label{eq: RG with c}
\frac{\partial Z}{\partial\xi}=-\frac{2}{\pi^2 N}\lp \xi +c\rp Z
% ,\quad
% c=\ln\frac{N^2\pi^2}8
% \xi=\ln \frac{E_0}{\Lambda}
,
\ee
where we retained a constant $c=\ln\frac{N^2\pi^2E_0}{8\Lambda_0}$ corresponding to the first term in the self energy renormalization, Eq.(\ref{eq: Z}). 

Integrating the RG equation, and taking into account the boundary conditions $Z(0) = \Gamma(0) = 1$, we obtain
%LL We can now apply the Ward identity Eq.(\ref{eq: Ward identity}) and 
%LL to solve the RG flow equations (\ref{eq: z}), (\ref{eq: Gamma}). We obtain 
\begin{equation}
Z(\xi) = \exp \left(-\frac{2c\xi+\xi^2}{N\pi^2}\right) = \Gamma^{-1}(\xi)
,\quad 
\xi=\ln \frac{\Lambda_0}{\Lambda}\label{eq: solution}.
\end{equation} 
We note that in the limit of small $\xi^2/N$, we reproduce the perturbative result\cite{Yang} for the residue, Eq.(\ref{eq: Z}). However, our result (\ref{eq: solution}) applies for all $\xi$, both small and large. The fermion propagator at arbitrary energies and momenta is then given by 
\begin{equation}
G(\omega, \vec{k}) = - Z(\xi) \frac{i\omega + H_0(\vec{k})}{\omega^2 + \big(\frac{\vec k^2}{2m}\big)^2}. \label{eq: final Green function}
\end{equation}  
At zero temperature, the infrared cutoff is supplied by the external frequency and momentum, 
%\mpar{changed $\ln \frac{N^2 \pi^2 E_0}{8 \Lambda}$ to $\ln E_0/\Lambda$?}
such that $\xi = \ln \frac{\Lambda_0}{\Lambda}$ and $\Lambda = \sqrt{\omega^2 + (k^2/2m)^2}$

Thus, the quasiparticle residue in undoped BLG is suppressed to zero by electron-electron interactions, Eq.(\ref{eq: final Green function}). This is reminiscent of the situation in disordered metals, where enhancement of interactions by disorder produces a renormalization of electron self energy of a $\log^2$ form \cite{Aronov}, and analysis of an RG flow\cite{Finkelstein} yields a suppression of the quasiparticle residue similar in form to our Eq.(\ref{eq: final Green function}). The suppression of the quasiparticle spectral weight at low energies, governed by the $Z(\xi)$ dependence, will manifest itself directly in the behavior of the tunneling density of states of BLG, similar to disordered metals.
% experiments, just as in disordered metals. In particular, as $Z$ flows to zero, the tunneling density of states will flow to zero also. 

We note parenthetically that, while keeping the constant term $c$ in the RG equation (\ref{eq: RG with c}) is formally beyond the $\log^2$ accuracy generally adopted in our analysis, it can be justified on the same grounds as in the discussion of the zero bias anomaly in disordered metals \cite{Nazarov,Levitov}. Because of its fairly large value for $N=4$, given by $c=\ln 2\pi^2\approx 2.98$, this term may significantly alter predictions for the behavior of $Z$ at intermediate energies $\epsilon\lesssim \Lambda_0$.

To analyze the suppression of tunneling density of states (TDOS), we use its relation to the retarded Green function \cite{Aronov},
%LL current more concrete, we note that the tunneling density of states $\rho(\omega)$ is given by \cite{Aronov}
%
\begin{equation}
\rho(\omega)= - \frac{1}{\pi}  \Im\left[ \Tr  G_R(\omega,\vec k)\right]
, \label{eq: spectral function}
\end{equation}
where $G_R(\omega,\vec k)$ is obtained from the Matsubara Green function analyzed above, Eq.(\ref{eq: final Green function}), by
%LL represents taking 
the analytic continuation of frequency from  imaginary to real values, $i\omega \rightarrow \omega + i \eta$.
%LL , and $G(\omega)$ is the Green function at Matsubara frequency $\omega$.

It is convenient to take the trace before performing the analytic continuation. The trace may be most easily taken in a basis of free particle eigenstates (plane waves with appropriate spinor structure), which amounts to integrating Eq.(\ref{eq: final Green function}) over all $\vec k$ values, $\Tr G= \int G(\omega,\vec k) d^2k$. Noting that the term containing $H_0(\vec k)$ vanishes upon integration due to the angular dependence, we write
\be
\Tr G= 
%G(\omega,\vec k) d^2k N\int G(\omega,\vec k) d^2k = 
\frac{2 N_0}{\pi}\int_0^\infty Z(\xi)\frac{i\omega}{\omega^2+z^2}dz
,
\ee
where $z = \vec k^2/2m$ and $N_0$ is the density of electronic states in BLG in the absence of interactions. 

It can be seen that the integral over $z$ is determined by $z\sim\omega$. It is therefore convenient to introduce a variable $\phi=\sinh^{-1}(z/\omega)$ and write
\be
\Tr G = i\frac{2 N_0}{\pi}\int_0^\infty Z(\xi_\omega-\ln\cosh\phi)\frac{d\phi}{\cosh\phi}
,
\ee
where $\xi_\omega=\ln\lp \Lambda_0/\omega\rp$. Noting that this integral is dominated by $\phi\sim 1$, we obtain an estimate of the spectral weight:
\be\label{eq: Zw}
\rho(\omega)\approx N_0 Z(\xi_\omega)=N_0\exp\lp - \frac{\xi_\omega^2+2c\xi_\omega}{N\pi^2}\rp
.
\ee
%
%\mpar{OK?}
The form of this expression remains unchanged, to leading $\log^2$ order, upon analytic continuation to real frequencies. 
%LL where we assume that the order of magnitude of $Z$ does not change upon analytic continuation to real frequencies. 
%\mpar{ref to appendix suppressed}
%LL A more careful calculation leading to the result Eq.\ref{eq: Zw} is provided in the appendix. 

\begin{figure}
\includegraphics[scale = 0.7]{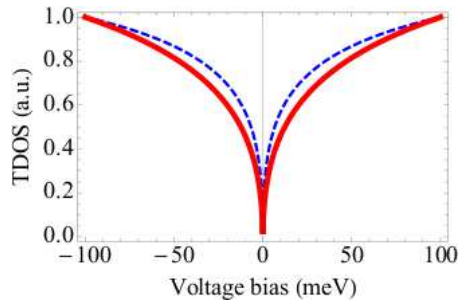} 
\caption{Tunneling density of states (TDOS) of BLG at charge neutrality, Eq.(\ref{eq: plottable}), is shown as a function of external bias $\omega = eV$. Predicted TDOS is shown for two different values of the dielectric constant in $E_0$, Eq.(\ref{eq: scales}): $\kappa=1$ (solid curve) and $\kappa=2.5$ (dashed curve), describing free-standing BLG and BLG on SiO substrate, respectively. Plot 
%LL goes over range $-100$meV$< \omega < 100 $meV, and 
is normalized so that $\rho = 1$ at an external bias of $100\,{\rm meV}$.}
\label{fig: anomaly}
\end{figure}

The expression in Eq.(\ref{eq: Zw}) can be re-arranged by using Eq.(\ref{eq: difference equation}) as 
\begin{equation}
\rho(\omega) = N_0 \exp \lp - \frac{\ln^2 \frac{N^2 \pi^2 E_0}{8 \omega} - \ln^2 \frac{N^2 \pi^2 E_0}{8 \Lambda_0}}{N \pi^2} \rp
.
\end{equation}
Thus, we see that the only effect of the UV cutoff $\Lambda_0$ is to 
%LL provide an overall 
rescale the prefactor for the TDOS
%LL tunneling density of states $\rho(\omega)$ 
without affecting the frequency dependence.
%LL - the functional dependence is completely independent of $\Lambda_0$, and we have \mpar{OK?}
Absorbing the dependence on $\Lambda_0$ in the prefactor, we have
\begin{equation}
\rho(\omega) = \tilde N_0 \exp \lp - \frac1{N \pi^2}\ln^2 \frac{N^2 \pi^2 E_0}{8 \omega} \rp
. \label{eq: plottable}
\end{equation} 
Tunneling measurements yield $\rho(\omega = eV)$, where $V$ is the bias voltage. 
The interaction suppression of the TDOS, Eq.(\ref{eq: Zw}), will therefore manifest itself as a zero bias anomaly in tunneling experiments. The predicted behavior the TDOS is shown in Fig.\ref{fig: anomaly}. 
%
%The suppression of the tunneling density of states predicted by Eq.(\ref{eq: Zw}) sets in around a hundredth of the Rydberg energy, Eq.(\ref{eq: scales}): for $\omega/E_0=0.1,\,0.01,\,0.001$, which lie well within the experimentally accessible range, we estimate the suppression factor 
%
%\be\label{eq: Z numbers}
%\rho(\omega)/N_0=Z(\xi_\omega)=0.62,\,0.3,\,0.1,
%\ee
%
Because of the exponential dependence in Eq.(\ref{eq: plottable}), the suppression rapidly becomes more pronounced at lower energies.

Closing our discussion of the zero bias anomaly in BLG, we note that the results described above apply only to the system at charge neutrality. Away from neutrality, with the Fermi surface size becoming finite, the effects of screening will grow stronger, resulting in a weaker effective interaction. Yet, even in this case, the tunneling density of states will be described by the suppression factor $\rho(\omega=eV)/N_0$ given by Eq.(\ref{eq: Zw}), provided that the bias voltage $eV$ exceeds the Fermi energy measured from the neutrality point.

%\mpar{text removed}

%Tunneling experiments measure $\rho(\omega = eV)$ \cite{Aronov}, where $V$ is the bias voltage. The interaction suppression of the density of states (Eq.(\ref{eq: answer})) therefore implies a zero bias anomaly in tunneling experiments on BLG. The predicted effect is of order one when 
%
%\begin{equation}
%\ln \frac{\Lambda}{\omega} = \sqrt{N} \pi \qquad \omega = \Lambda \exp(-\sqrt{N}\pi) \label{eq: bandwidth}
%\end{equation}
%
%Within the weak coupling approximation, $\Lambda = N^2 E_0$. Upon making an analytic continuation from large $N$ and small $E_0$ to $N=4$ and $E_0$ defined by Eq.(\ref{eq: scales}), we have $\Lambda \approx 24 eV \kappa^{-2}$. This suggests that the zero bias anomaly sets in at $\omega \approx 45 meV$. In practice, however, a bandwidth of $24 eV \kappa^{-2}$ is certainly an overestimate. The bandwidth of the $\pi$ bands in graphene is $12 eV$, and this sets the upper limit for the bandwidth. Also, on energy scales greater than $1eV$, mixing with the upper bands of BLG becomes significant, and a four band analysis becomes necessary. Therefore, it is likely that in practice the bandwidth $\Lambda$ in Eq.(\ref{eq: bandwidth}) is no more than a few $eV$. However, even a conservative estimate of a few $eV$ for the bandwidth yields an energy scale of several $meV$ as the threshold for the zero bias anomaly, and this lies well within the experimentally accessible range. 

\section{Single log renormalization of electron mass}
\label{sec: mass}
Thus far we have concentrated on $\log^2$ flows. However, the analysis may be extended to obtain the subleading single log flows of the action. We illustrate this procedure by calculating the renormalization of the mass (which did not renormalize at $\log^2$ order in the RG). This calculation is  interesting because it allows us to 
%LL predict 
investigate the interaction renormalization of the compressibility--a directly measurable quantity, and also because it allows us to illustrate how much slower the single $\log$ flows are than the $\log^2$ flows. 

In this section, we first analyze mass renormalization by extracting it directly from the self energy. After that, in Sec.\ref{sec: compressibility} we consider electron compressibility of BLG and show that the $\log$ divergent correction to the compressibility matches exactly our prediction for mass renormalization obtained from the self energy. 

In BLG, the self energy is a $2\times 2$ matrix, given by Eq.(\ref{eq: self energy})), which is related to the renormalized Green function by the Dyson equation, 
%LL It is convenient, instead of considering just the leading order renormalization of the propagator, to resum an infinite series of one-particle irreducible diagrams, to obtain
%
\begin{equation}
G^{-1}(\omega, \vec{q}) = G_0^{-1}(\omega, \vec{q}) - 
\lp
\begin{array}{cc}
\Sigma_{AA}(\omega, \vec{q}) &\Sigma_{AB}(\omega, \vec{q}) \\ \Sigma_{BA}(\omega, \vec{q}) &\Sigma_{BB}(\omega, \vec{q})
\end{array}
\rp
.
%LL \Sigma(\omega, \vec{q}) 
\label{eq: mass}
\end{equation}
%
%LL This is identical to Eq.(\ref{eq: self energy}) for an infinitesimal RG transformation, when $\Sigma \sim d \xi$. 
As discussed in Sec.\ref{sec4}, the leading $\log^2$ contribution to the self energy is proportional to $G_0^{-1}$, since $\partial \Sigma_{AB}/\partial (q_+^2/2m)= i\partial \Sigma_{AA}/\partial \omega$. This means that all renormalization can be attributed to the residue $Z$ with mass remaining unchanged. However, as we now show, this equality is only true to leading logarithmic order. 

Comparison of Eq.(\ref{eq: mass}) with Eq.(\ref{eq: sigmaaa}) and Eq.(\ref{eq: sigmaab}) 
%LL tells us 
indicates that the mass renormalization is given by
\begin{equation}
\frac{\delta m}{m} = Z_0 \lp i\frac{\partial \Sigma_{AA}}{\partial \omega} - \frac{\partial \Sigma_{AB}}{\partial (q_+^2/2m)} \rp
.  \label{eq: mass recursion}
\end{equation}
Here, $i\partial \Sigma_{AA}/\partial \omega$ is defined by Eq.(\ref{eq: res renormalization}). For the second term, we obtain the expression
%
%\begin{widetext}
\begin{eqnarray}
&& \frac{\partial \Sigma_{AB}}{\partial (q_+^2/2m)} = \int \frac{d\epsilon d^2p}{(2\pi)^3} \left(\frac{1}{\epsilon^2 + \big(\frac{p^2}{2m}\big)^2} - \frac{5 \big(\frac{p^2}{2m}\big)^2}{\lp \epsilon^2 + \big(\frac{p^2}{2m}\big)^2\rp ^2}  \right.
\nonumber \\
&& \left. + \frac{4 \big(\frac{p^2}{2m}\big)^4}{\lp \epsilon^2 + \big(\frac{p^2}{2m}\big)^2\rp ^3} \right) \Gamma^2 Z U(\epsilon, \vec{p})
, \label{eq: dqsquared}
\end{eqnarray}
%\end{widetext}
where $U(\epsilon, \vec{p})$ is given by Eq.(\ref{eq: Ueff}). To evaluate the difference in Eq.(\ref{eq: mass recursion}), 
%LL $i\partial \Sigma_{AA}/\partial \omega - \partial \Sigma_{AB} / \partial (q_+^2/2m)$, 
it is convenient to subtract the integrands of Eqs.((\ref{eq: res renormalization},\ref{eq: dqsquared})) before doing the integrals. Once again, we 
%LL express momenta in polar coordinates, $p_x = p \cos \zeta$, $p_y = p \sin \zeta$, and straightaway integrate over $\zeta$. After changing to 
use the ``polar'' representation of the frequency and momentum variables, $\omega = r \cos \theta$, $p^2/2m = r \sin \theta$, and obtain
\begin{equation}
\frac{\delta m}{m}= \int_{\Lambda'}^{\Lambda} \frac{dr}{r} \int_{0}^{\pi} \frac{d\theta}{2\pi} \frac{\Gamma_0^2 Z_0^2 (3\sin^2\theta - 4\sin^4 \theta)}{\sqrt{2r\sin \theta} - \frac{2 \pi}{m} \Pi(\theta)} ,\nonumber
\end{equation}
where $\Pi(\theta)$ is the polarization function introduced in Eq.(21), and $r$ is measured in units of $E_0$ as before.
%LL We are measuring $r$ as always in units of $E_0$. 
The integral over $\theta$ is now fully convergent, and the resulting expression is only single log divergent. Integrating analytically over $r$ and then integrating numerically over $\theta$, we find
\begin{equation}
\frac{\delta m}{m} = \frac{0.56  }{2N\pi \ln 4} \Gamma_0^2 Z_0^2 \ln \frac{\Lambda}{\Lambda'}
.
\end{equation}
Converting this recursion relation into a differential equation, we obtain
\begin{equation}
\frac{d \ln m}{d\xi} = \frac{0.56 }{2N \pi \ln4}\Gamma^2 Z^2 
. \label{eq: mass flow}
\end{equation}
This equation cannot be solved for general $\xi$ by applying the Ward identity Eq.(\ref{eq: Ward identity}), since the Ward identity only holds at leading $\log^2$ order, while the mass flows at subleading (single log) order in Eq.(\ref{eq: mass flow}). In the perturbative limit $\frac1{N}\xi \ll 1$, when $Z\approx 1$ and $\Gamma \approx 1$, we obtain a logarithmic correction to the mass 
\begin{equation}
m (\xi)= m(0) \lp 1 + \frac{0.56}{2N \pi \ln 4} \xi\rp 
. 
\label{eq: mass solution}
\end{equation}
We may relate 
this mass renormalization to a measurable quantity, by noting that the electronic compressibility $K$ is proportional to the density of states which is proportional to the mass. Thus, the logarithmic renormalization of the mass in Eq.(\ref{eq: mass solution}) should manifest itself in a logarithmic enhancement of the electronic compressibility. The relation between mass renormalization and compressibility will be further discussed in Sec.\ref{sec: compressibility}.

%To obtain the flow equation for arbitrary $\xi$, not just $\xi \approx 1$, we go through the same procedure that was followed for $Z$. By comparing Eq.\ref{eq: Green function} with Eq.\ref{eq: renormalizations}, and projecting the self energy $\Sigma$ on the conduction band to avoid having to deal with matrices, we get
%
%\begin{equation}
%\frac{\delta \Sigma_{c}(\xi)}{\delta \xi} = \frac{1}{Z^2(\xi)} \frac{\delta Z(\xi)}{\delta \xi} \left(i\omega - \frac{p^2}{2m(\xi)}\right) - \frac{p^2}{2m^2(\xi)Z(\xi)} \frac{\delta m(\xi)}{\delta \xi} \label{eq: m1}
%\end{equation}

%Meanwhile, we can also write 
%\begin{equation}
%\Sigma_c(\xi) = -i \frac{\partial \Sigma_c}{\partial \omega} \left(i \omega - \frac{p^2}{2m(\xi)}\right) - \frac{p^2}{2m(\xi)} \left(i\frac{\partial \Sigma_c}{\partial \omega} - \frac{\partial \Sigma_c}{\partial (p^2/2m)}\right)
%\end{equation}
%where 
%\begin{equation}
%\left(i\frac{\partial \Sigma_c}{\partial \omega} - \frac{\partial \Sigma_c}{\partial (p^2/2m)}\right)
%= \Gamma^2(\xi) Z(\xi) \frac{0.56 \ln \xi}{2N\pi \ln4} \label{eq: m2}
%\end{equation}
%Comparing Eq.\ref{eq: m1} with Eq.\ref{eq: m2}, we get 
%\begin{equation}
%\frac{\partial m}{\partial \ln \xi} = \Gamma^2(\xi) Z^2(\xi) m(\xi) \frac{0.56}{2N \pi \ln 4}
%\end{equation}
%Solving this equation, again using $\Gamma Z = 1$, we obtain 
%\begin{equation}
%m(\xi) = m(1) (\xi)^{0.56/(2N\pi \ln 4} \label{eq: mscaling}
%\end{equation}

\section{Interaction correction to compressibility}
\label{sec: compressibility}

%LL We can formalize this argument by explicitly calculating the interaction 
Here we explicitly calculate the
renormalization of the compressibility. 
By doing this we shall confirm that the compressibility does not renormalize at leading (log square) order, and also extract the single log renormalization of the compressibility.
The interaction correction to the compressibility $K$ is given by 
\be\label{eq: K=d2F}
\delta K = - \frac{\partial^2 F}{\partial \mu^2}
,
\ee
where $\mu$ is the chemical potential, and $F$ is the interaction energy. Within the RPA framework, the interaction energy is expressed as 
\begin{equation}
F(\mu) = \int \frac{d\omega d^2 p}{(2\pi)^3} \ln\left(1 - V(\vec{q}) \Pi(\mu, \omega, \vec{q})\right)
. \label{eq: RPA free energy}
\end{equation}
Here, $\Pi(\mu, \omega, \vec{q})$ is the non-interacting polarization function evaluated at a chemical potential $\mu$, and $V(q)$ is the unscreened Coulomb interaction $V(q) = 2\pi e^2/\kappa q$. 

To evaluate the second derivative in (\ref{eq: K=d2F}), we consider the difference $\Delta F=F(\mu)-F(0)$. After rearranging logs under the integral, we rewrite this expression as
%LL We now wish to calculate the interaction correction to the electronic compressibility, to confirm that the compressibility does not renormalize at leading (log square) order, and also to extract the single log renormalization of the compressibility. We start with the (infinite order) RPA expression for the interaction energy as a function of chemical potential $F(\mu)$, Eq.\ref{eq: RPA free energy}. We then rearrange it to obtain
%
\begin{equation}
%LL F(\mu) - F(0) 
\Delta F= - \int \frac{d\omega d^2 q}{(2\pi)^3} \ln \bigg(1 - U_{\omega, q} \big(\Pi(\mu, \omega, q) - \Pi(0, \omega, q)\big)\bigg), 
\label{eq: RPA free energy 2}
\end{equation}
where now $U_{\omega, q}$ is the dynamically screened Coulomb interaction, Eq.(\ref{eq: Ueff}). Since the compressibility is obtained from the free energy through $K = - \partial^2 F/\partial \mu^2$, the problem of calculating the interaction renormalization of the compressibility is reduced to that of calculating the polarization function at finite $\mu$. This may be calculated through methods similar to those developed in Ref.\onlinecite{Nandkishore}. We define $\epsilon_{\pm} = \epsilon \pm \omega/2$, $\vec{p}_{\pm} = \vec{p} \pm \vec{q}/2$ and $z_{\pm} = |\vec{p}_{\pm}|^2/2m$. The non-interacting polarization function at finite $\mu$ is given by 
\begin{eqnarray}
&&\Pi(\mu, \omega, q) = \Tr  G(\mu, \epsilon_+, \vec{p}_+) G(\mu, \epsilon_-, \vec{p}_-) \nonumber \\
&&= \Tr  \int \frac{d\epsilon d^2p}{(2\pi)^3} \frac{1}{\big(i\epsilon_+ - \mu - H_0(\vec{p}_+)\big)\big(i\epsilon_- - \mu - H_0(\vec{p}_-)\big)} \nonumber\\
%&&= - 2 \int \frac{d\epsilon d^2p}{(2\pi)^3\big(\epsilon_+^2 - \mu^2 + z_+^2 + 2i\epsilon_+ \mu\big)}\nonumber\\
%&&\phantom{AAAAA}\times \frac{\epsilon_+ \epsilon_- - z_+z_- \cos 2 \theta_{pq} - \mu^2 + 2i \mu \epsilon}{\big(\epsilon_-^2 - \mu^2 + z_-^2 - 2i\epsilon_- \mu\big)}
%\nonumber\\
&&= 2N \int \frac{d\epsilon d^2p}{(2\pi)^3 \big(\epsilon_+ + i(\mu + z_+)\big)\big(\epsilon_+ + i(\mu - z_+)\big)} \nonumber\\
&&\phantom{AAAAA}\times \frac{(i\epsilon_+-\mu)(i\epsilon_- - \mu) + z_+z_-\cos 2\theta_{pq}}{\big(\epsilon_- + i(\mu +z_-)\big)\big(\epsilon_- + i(\mu - z_-)\big)}
, \label{eq: mupolfn}
\end{eqnarray}
where $\theta_{pq}$ is the angle between $\vec{p}_+$ and $\vec{p}_-$. We now perform the integral over $\epsilon$ by residues to obtain
%
% \begin{widetext}
\begin{eqnarray}
&& \Pi(\mu, \omega, \vec{q}) = N\int \frac{d^2p}{(2\pi)^2} \frac{\big(z_+ + i\omega + z_- \cos 2\theta_{pq}\big) \Theta(z_+-\mu)}{z_+^2 - z_-^2 - \omega^2 + 2i\omega z_+}  \nonumber\\
&& \phantom{AAAAA} + \lp \omega, \vec{q} \rightarrow - \omega, - \vec{q}\rp\nonumber\\
&& = N\int_{z_+ = 0}^{z_+ = \mu} \frac{d^2 p_+}{(2\pi)^2} \lb \frac{1}{z_+ + i\omega - z_-} - \frac{2z_- \sin^2 \theta_{pq}}{(z_+ + i\omega)^2 - z_-^2}\rb \nonumber\\
&& \phantom{AAAAA} + \lp \omega, \vec{q} \rightarrow - \omega, - \vec{q}\rp 
. \label{eq: mupolfn2}
\end{eqnarray}
%
% \end{widetext}
In the limit $\mu \rightarrow 0$, this reproduces the non-interacting polarization function from Ref.[\onlinecite{Nandkishore}]. Now we expand Eq.(\ref{eq: RPA free energy 2}) to leading order in small $\mu$ to obtain
\begin{equation}
%LL F(\mu) - F (0) 
\Delta F = - \frac{1}{2}\mu^2 \int \frac{d\omega d^2 q}{(2\pi)^3} U(\omega, q) \frac{\partial^2\Pi(\mu, \omega, q)}{\partial \mu^2}
. \label{eq: RPA free energy 3}
\end{equation}
The term linear in $\mu$ must vanish, by particle hole symmetry. Taking derivatives of Eq.(\ref{eq: mupolfn2})
%LL , meanwhile, 
greatly simplifies the calculations, since it turns the two dimensional integral over momenta into a one dimensional integral over momentum angles, which is fully convergent, and may be evaluated numerically. We find 
%$\partial \Pi/\partial \mu = 0$ (as expected on symmetry grounds), and 
%
\begin{eqnarray}
&&\frac{\partial^2 \Pi}{\partial \mu^2} = \frac{Nm}{2\pi} \frac{3 \omega^2 z_q^2 - z_q^4}{(\omega^2 + z_q^2)^2}, \qquad z_q = \frac{q^2}{2m}
, \\
&& \Delta F = - \frac{\mu^2}{2} \int \frac{d\omega d^2q}{(2\pi)^3} U(\omega, \vec{q}) \frac{\partial^2 \Pi}{\partial \mu^2}
.
\label{eq:Fm-F0}
\end{eqnarray}
%
%where $z_q = q^2/2m$. \mpar{no integral?}
We again
%LL now 
change to the coordinates $\omega = r \cos \theta$, $z_q = r \sin \theta$, and measure $r$ in units of $E_0$. Note that even though the interaction has a pole at $\theta \rightarrow 0, \pi$, this pole is canceled by $\partial^2 \Pi/\partial \mu^2$ having a zero at $\theta \rightarrow 0, \pi$. As a result, the $\theta$ integral is fully convergent. Integrating numerically over $\theta$ and analytically over $r$, we find that the fractional change in the compressibility is 
\begin{equation}
\frac{\delta K(\xi) }{K(0)}=  \frac{0.56}{2N \pi \ln 4} \xi 
,
\end{equation}
a result that agrees exactly with Eq.(\ref{eq: mass solution}). We note that an
%LL the predicted interaction 
enhancement of the compressibility due to interactions was also predicted in 
%LL is in agreement with 
Ref.[\onlinecite{Kusminsky}]. 
%\mpar{OK?}
However, the effect described by
%LL enhancement 
Eq.(\ref{eq: mass solution}) is much weaker than that predicted in Ref.[\onlinecite{Kusminsky}], because we have worked with a screened interaction, whereas in Ref.[\onlinecite{Kusminsky}] screening was not taken into account.

%LL We note that 
In summary, the compressibility does not renormalize at leading (log square) order, just as in the Luttinger liquids, and while there is a subleading logarithmic correction, the pre-factor is quite small ($0.56/(2N \pi \ln 4)\approx 0.016$). Thus, in contrast to the zero-bias anomaly in TDOS, experimental detection of the interaction correction to the compressibility is likely to be challenging.
%LL prove difficult to detect. \mpar{OK?} The contrast with the strong zero bias anomaly in the tunneling density of states is clear. 
The difference arises because the single log flows are much weaker than the $\log^2$ flows, retrospectively justifying our earlier neglect of the single log flows in the RG. 
% \mpar{OK?}
Hence, strong suppression of the tunneling density of states at energy scales where the compressibility is not significantly renormalized is a key signature of the marginal Fermi liquid physics in bilayer graphene.

\section{Discussion and Conclusions}

%LL We comment now on the regime 
Here we briefly discuss the range of validity of our results. Our analysis was organized as a perturbation theory in $\Gamma^2Z^2 / N$. Since $\Gamma Z=1$ at leading (log square) order, the perturbation theory remains well defined under the log square flows. However, our analysis neglected subleading single log flows. For $\xi \approx N \pi^2$, the subleading single log flows become important, and the analysis leading to the expression Eq.(\ref{eq: solution}) no longer applies. A mean field theory of subleading single log effects \cite{Nandkishore} indicates that a gapped state develops at $\xi = \frac{3}{13} N \pi^2$, the scale which we tentatively identify as the limit of validity of our analysis. 

How can the marginal Fermi liquid physics be distinguished from the formation of a gapped state? We note that at very low energies, once the gapped state has developed, the tunneling density of states will vanish anyway. However, in the gapped regime, the compressibility will vanish also. What we have shown, however, is that there is a large range of energies
%LL energy scales (temperatures) 
greater than the 
%LL critical 
energy scale for gap formation, where the tunneling density of states vanishes, while the compressibility remains essentially unchanged. Such behavior represents the key signature of the marginal Fermi liquid physics discussed above, which is analogous to the Luttinger liquid physics.
%LL lies therefore in the vanishing of the tunneling density of states at an energy scale (temperature) where the compressibility is essentially the same as the non-interacting value. 

%LL We have also 
In our analysis, we neglected the short range interactions which are characterized by lattice scale, such as the interlayer density difference interaction $V_- =\frac12 (V_{AA} -V_{AB})= \pi e^2 d$ and the Hubbard-type on-site repulsion. 
%\mpar{OK?}
Short range interactions are non-dispersive, do not renormalize the Green function in the weak coupling limit, and hence do not alter our results. Short range interactions also produce only single log renormalization \cite{Zhang, Vafek} and therefore do not need to be included in our log square RG. %Lattice scale interactions will, however, be important in the subleading RG of single log flows, where they will control the nature of the phase that is realized at very low energies. The RG of single log flows will be discussed in a future publication. %since they renormalize logarithmically under RG, and can produce strong coupling instabilities  \cite{Zhang, Kivelson, Vafek}. However, these strong coupling instabilities are single log instabilities, which develop only on an energy scale $\xi \approx 1/g$  \cite{Zhang, Kivelson, Vafek}, where the bare short range interaction $g$ is assumed to be small. Thus, short range interactions provide an alternative infrared cutoff for our RG, $\xi_{max} = \max(\frac{3}{13} N \pi^2, \frac{1}{g}, T)$, but they do not alter our log \emph{squared} RG. 
Similarly, we justify our neglect of the trigonal warping effect \cite{McCannFalko} by noting that trigonal warping is significant only on energy scales smaller than the characteristic energy scale for onset of gapped states \cite{Nandkishore}. 

Finally, we note that our analysis made use of the fact that there were no un-canceled log square divergences at one loop order in the RG, except for the renormalization of the quasiparticle residue and the Coulomb vertex function, which were related by a Ward identity, Eq.(\ref{eq: Ward identity}). Technically, in order for our neglect of higher loop corrections to be justified, we also require that there are no un-canceled log square divergences beyond one loop order in the RG, except those that are constrained by Ward identities. We believe this to be the case, however, the proof requires a non-perturbative approach, which lies beyond the scope of the present work. 

To conclude, we have examined the one-loop RG flow for bilayer graphene. We have demonstrated that the quasiparticle residue $Z$ and the Coulomb vertex function $\Gamma$ both flow as $\xi^2$, where $\xi$ is the RG time. All other quantities flow only as $\xi$. 
%LL We have self-consistently solved the RG equations with logarithmic accuracy. 
The structure of the RG for Coulomb interacting BLG has strong similarities to the RG for the one dimensional Luttinger liquids. In particular, we predict a strong interaction suppression of the tunneling density of states for undoped BLG, even at energy scales where the electronic compressibility is essentially unchanged from its non-interacting value. These predictions may be readily tested by experiments. 
%LL If a zero bias anomaly in the tunneling density of states is observed, it would constitute a striking observation of marginal Fermi liquid behavior in bilayer graphene. 
%LL Coulomb interacting bilayer graphene. 

%We note that our analysis was developed for charge neutral BLG. For small doping $\mu \ll \Lambda$ away from charge neutrality, the above formalism will apply, with $\mu$ acting as an infrared cutoff.  For large doping, the system may be viewed as a two dimensional chiral electron gas, which is not expected to display a zero bias anomaly. Technically, the special form of the screened interaction (Eq.\ref{eq: Ueff}) leading to $\log^2$ divergences is particular to charge neutral BLG - well away from charge neutrality the analysis developed here does not apply. 

We acknowledge useful conversations with A. Potter and P. A. Lee. This work was supported by Office of Naval Research Grant No. N00014-09-1-0724.

\end{document}